\documentclass[aps,prd,preprint,superscriptaddress,nofootinbib]{revtex4-2}

\usepackage[T1]{fontenc}    
\usepackage[utf8]{inputenc} 
\usepackage{lmodern}        

\usepackage{amsmath,amssymb,amsthm,amsfonts} 
\usepackage{bm}             
\usepackage{slashed}        
\usepackage{physics}        

\usepackage{graphicx}       
\usepackage[caption=false]{subfig} 
\usepackage{booktabs}       
\usepackage{multirow}       
\usepackage{orcidlink}
\usepackage{simpler-wick}
\usepackage[dvipsnames]{xcolor} 
\usepackage{hyperref}       
\hypersetup{
    colorlinks=true,
    linkcolor=blue,
    citecolor=blue,
    urlcolor=blue
}

\allowdisplaybreaks

\begin{document}


\preprint{}

\title{The $\Omega(2380)$ as a partner of the $\Omega(2012)$}

\author{Yi-Yao Li\orcidlink{0009-0001-6943-4646}}
\email{liyiyao@m.scnu.edu.cn}
\affiliation{
State Key Laboratory of Nuclear Physics and 
Technology, Institute of Quantum Matter, South China Normal 
University, Guangzhou 510006, China}
\affiliation{Key Laboratory of Atomic and Subatomic Structure and Quantum Control (MOE), Guangdong-Hong Kong Joint Laboratory of Quantum Matter, Guangzhou 510006, China }
\affiliation{ Guangdong Basic Research Center of Excellence for Structure and Fundamental Interactions of Matter, Guangdong Provincial Key Laboratory of Nuclear Science, Guangzhou 510006, China  }
\affiliation{Departamento de Física Teórica and IFIC, Centro Mixto Universidad de Valencia-CSIC, Institutos de Investigación de Paterna, 46071 Valencia, Spain}

\author{Albert Feijoo\,\orcidlink{0000-0002-8580-802X}}
\email{edfeijoo@ific.uv.es}
\affiliation{ Instituto de Física Corpuscular, Centro Mixto Universidad de Valencia-CSIC, Institutos de Investigación de Paterna, Aptdo. 22085, 46071 Valencia, Spain}

\author{Eulogio Oset\,\orcidlink{0000-0002-4462-7919}}
\email{eulogio.oset@ific.uv.es}
\affiliation{Departamento de Física Teórica and IFIC, Centro Mixto Universidad de Valencia-CSIC, Institutos de Investigación de Paterna, 46071 Valencia, Spain}
\affiliation{\small Department of Physics, Guangxi Normal University, Guilin 541004, China}

\begin{abstract}
We present a study of the $\Omega(2380)$ resonance and show that it is consistent with a dynamically generated state arising from the $\bar{K}^*\Xi^*$, $\omega\Omega$, and $\phi\Omega$ interactions. In this picture, the $\Omega(2380)$ is analogous to the $\Omega(2012)$, which is generated from the $\bar{K}\Xi^*$ and $\eta\Omega$ channels. The resulting mass, total width, and partial decay widths into the $\bar{K}\Xi^*$ and $\bar{K}^*\Xi$ channels are compatible with the available experimental data. We also discuss possible experimental observables that could provide further insight into the nature of this state.
\end{abstract}

\maketitle

\section{Introduction}
\label{Sec:intro}

The spectroscopy of excited $\Omega^*$ baryons has long remained poorly explored experimentally. For nearly four decades, only three states, i.e. $\Omega(2250)$, $\Omega(2380)$ and $\Omega(2470)$, with two- or three-star status were listed in the Review of Particle Physics (RPP) \cite{ParticleDataGroup:2018ovx}. This limited experimental information stood in stark contrast to a wide range of theoretical studies based on quark models, Skyrme-type approaches, and lattice QCD (LQCD), which anticipated a significantly richer $\Omega^*$ spectrum
\cite{Capstick:1986ter,Loring:2001ky,Pervin:2007wa,Faustov:2015eba,Chao:1980em,Kalman:1982ut,An:2013zoa,An:2014lga,Oh:2007cr,Engel:2013ig,CLQCD:2015bgi}.

A major breakthrough came with the observation of a narrow resonance, the $\Omega(2012)^-$, by the Belle Collaboration in $e^+e^-$ collisions \cite{Belle:2018mqs}. The state was reconstructed in the $K^0_S\Xi^-$ and $K^-\Xi^0$ decay channels and exhibited a width of about $6$~MeV. Comparisons with LQCD calculations \cite{Engel:2013ig} and conventional quark-model (CQM) predictions
\cite{Capstick:1986ter,Loring:2001ky,Oh:2007cr,Pervin:2007wa,Faustov:2015eba} favored a spin--parity assignment of $J^P=3/2^-$, consistent with a $d$-wave decay, whereas an $s$-wave $J^P=1/2^-$ assignment would imply a substantially larger width. Subsequent Belle analyses \cite{Belle:2019zco,Belle:2022mrg} refined the decay properties of the $\Omega(2012)^-$ and provided a precise determination of the branching-fraction ratio $\mathcal{R}^{\Xi\bar{K}\pi}_{\Xi\bar{K}}$ \cite{Belle:2022mrg}, which serves as a stringent constraint on theoretical interpretations. Independent evidence for $\Omega(2012)^-$ production was later reported by the BESIII Collaboration in the process $e^+e^- \to \Omega(2012)^- \bar{\Omega}^+ + c.c.$, together with indications of a new structure, the $\Omega(2109)^-$ \cite{BESIII:2024eqk}. More recently, the ALICE Collaboration observed a pronounced signal near $2013$~MeV in the $K^0_S\Xi^-$ invariant-mass distribution in $pp$ collisions at $\sqrt{s}=13$ TeV, providing a third independent confirmation of the $\Omega(2012)^-$ resonance \cite{ALICE:2025atb}. Beyond establishing its existence, these results provide a valuable benchmark for testing theoretical approaches to excited $\Omega$ states, particularly those based on unitarized meson--baryon dynamics.

The growing body of experimental information has triggered extensive theoretical investigations into the nature of the $\Omega(2012)^-$. Numerous studies based on constituent quark models interpret the state as a $p$-wave excited baryon
\cite{Xiao:2018pwe,Aliev:2018syi,Aliev:2018yjo,Wang:2018hmi,Polyakov:2018mow,Liu:2019wdr,Liu:2020yen,Arifi:2022ntc,Zhong:2022cjx,Wang:2022zja,Luo:2025cqs}. In contrast, alternative approaches emphasize the proximity of the $\Omega(2012)^-$ mass to the $\bar{K}\Xi(1530)$ threshold and describe it as a dynamically generated hadronic molecule within coupled-channel frameworks
\cite{Valderrama:2018bmv,Lin:2018nqd,Pavao:2018xub,Huang:2018wth,Gutsche:2019eoh,Lu:2020ste,Ikeno:2020vqv,Lin:2019tex,Ikeno:2022jpe,Lu:2022puv,Han:2025gkp,Shen:2025xcq}. It is worth noting that the revised Belle measurement of the branching-fraction ratio \cite{Belle:2022mrg} provides strong support for the molecular interpretation.

The use of unitarized effective theories (UEFTs) in coupled channels to describe meson--baryon interactions in the strangeness \(S=-3\) sector dates back more than two decades. The dynamical generation of \(S=-3\) baryon resonances from systems composed of pseudoscalar mesons and ground-state decuplet baryons (\(J^P = 3/2^+\)) was investigated in Refs.~\cite{Kolomeitsev:2003kt,Sarkar:2004jh,Xu:2015bpl}. In addition, the \(\bar{K}\Xi\) interaction was studied within an extended chiral \(SU(3)\) quark model by solving a resonating group method equation in Ref.~\cite{Wang:2008zzz}. More recently, the exploratory study of Ref.~\cite{Feijoo:2024qgq} examined the possibility of generating \(\Omega^*\) or \(P_{sss}\) (pentaquark) states from the \(\bar{K}\Xi\) interaction through attractive next-to-leading-order corrections to the chiral Lagrangian. A particularly relevant contribution in this theoretical context is Ref.~\cite{Gamermann:2011mq}, where a consistent \(SU(6)\) extension of the meson--baryon chiral Lagrangian is implemented within a coupled-channel unitary approach. In that work, a wide set of meson--baryon channels is considered, including all possible combinations of pseudoscalar and vector mesons with ground-state octet and decuplet baryons. As a result, nine \(\Omega^*\) resonances are predicted, two of which are compatible with the experimentally observed \(\Omega(2250)\) and \(\Omega(2380)\). However, the state associated with the \(\Omega(2380)\) resonance in that work appears as a bound state with a mass of about \(1930\)~MeV, predominantly coupling to the \(\bar{K}\Xi^*\), \(\bar{K}^*\Xi\), and \(\eta\Omega\) channels, and with non-negligible couplings to \(\bar{K}^*\Xi^*\) and \(\phi\Omega\). Such a low mass automatically precludes its decay into the \(\bar{K}\Xi^*\) and \(\bar{K}^*\Xi\) channels, thereby conflicting with the experimental information on the \(\Omega(2380)\) resonance~\cite{ParticleDataGroup:2024cfk}. This sizable mass discrepancy highlights a persistent difficulty of current coupled-channel approaches including vector mesons and decuplet baryons in accommodating the 
$\Omega(2380)$ within a realistic dynamical framework.

Despite these advances, a unified description of the $\Omega(2380)$ resonance within unitarized effective field theories remains unresolved. In particular, existing coupled-channel approaches either predict states significantly below the experimental mass or fail to reproduce simultaneously the observed decay modes and the sizable width of the $\Omega(2380)$, indicating that its dominant dynamical components are still poorly understood. It remains unclear which combinations of vector–decuplet and other meson–baryon channels might generate the observed resonance properties, motivating a detailed investigation in the present work.

In this context, in Ref.~\cite{Sarkar:2010saz}, the interaction between the vector-meson octet and the baryon decuplet was investigated using Lagrangians derived from the local hidden gauge symmetry approach. In that study, the unitary coupled-channel amplitudes in the \(S=-3\), \(I=0\) sector develop a pole that the authors associated with the \(\Omega(2470)\) resonance, given the proximity between the predicted mass of the state (\(2449\)~MeV) and the experimentally reported value. Although this assignment is plausible due to the absence of mixing with lighter channels---which would be expected to increase the width of the state---it nonetheless exhibits a significant discrepancy between the predicted and experimental widths, with the latter being approximately an order of magnitude larger than the theoretical estimate.

Interestingly, both the quark model developed in Ref.~\cite{Engel:2013ig} and the lattice-QCD-based studies of Refs.~\cite{CLQCD:2015bgi} indicate a direct association between pure $|sss\rangle$ configurations and the experimentally observed $\Omega(2470)$ state, while no clear three-quark configuration can be identified with the $\Omega(2380)$ resonance. This pattern has been largely sustained in subsequent theoretical investigations. This absence of a clear three-quark counterpart for the $\Omega(2380)$ strongly suggests that non-$sss$ components may play a dominant role in its structure.

Bearing in mind the elusive $|sss\rangle$ nature of the $\Omega(2380)$, the proximity of the $\bar{K}^*\Xi^*$ threshold to its mass, and the success of the molecular approach in identifying the dominant meson--baryon nature of the $\Omega(2012)$ state, it is timely to reexamine the theoretical $\Omega$-spectrum within the framework of unitarized effective field theories. In particular, this motivates a renewed study of the interaction between vector mesons and the baryon members of the decuplet. Though within the framework of the quark delocalization color screening model, the interaction between $\Omega$ baryon and $s\bar{s}$ mesons (i.e. $\phi$ and $\eta'$) has been addressed very recently \cite{Yan:2026yrd}, and although the $\Omega \phi$ interaction in the $J^P=1/2^-$ component was found to be attractive,  the corresponding strength was not sufficient to bind the system or generate a resonant state.  

In the present work, to assess the molecular nature of the $\Omega(2380)$ resonance, we revisit the vector meson-decuplet baryon interaction following Ref.~\cite{Sarkar:2010saz}, with the explicit inclusion of the $\Xi(1530)^0K^-$ and $\Xi^-\bar{K}^{*0}$ channels via box diagrams—corresponding to the only experimentally observed two-body decay modes of the $\Omega(2380)$. This allows for a more phenomenologically consistent estimation of the resonance width.

\section{Formalism}
\label{Sec:Formalism}
The work of Ref.~\cite{Pavao:2018xub} studied the $\Omega(2012)$ state in the $\bar{K}\Xi^*$, $\eta\Omega$ coupled channels, with $\bar{K}\Xi$ as the main decay mode. But in the present work, we turn to vector meson-decuplet baryon sector to study the interaction and the possible generation of bound states. The corresponding channels are  $\bar{K}^*\Xi^*(1530)$, $\omega\Omega$, $\phi\Omega$.

\begin{figure}[h!]  
    \centering
    \includegraphics[width=0.5\textwidth]{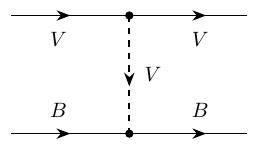} 
    \caption{Diagrammatic representation of the $VB\to VB$ interaction, through the exchange of vector meson.}
    \label{fig:MB}
\end{figure}

Fig.~\ref{fig:MB} depicts the interaction of meson-baryon via vector meson exchange following the local hidden gauge approach. The upper $VVV$ vertex is driven by the Lagrangian~\cite{Bando:1984ej,Bando:1987br,Meissner:1987ge,Nagahiro:2008cv}:
\begin{align}
    \mathcal{L}_{\mathrm{VVV}} &= i g \left\langle \left( V^{\mu} \partial_{\nu} V_{\mu} - \partial_{\nu} V^{\mu} V_{\mu} \right) V^{\nu} \right\rangle, \label{eq:VVV}
\end{align}
with the coupling constant  
$g = \frac{m_V}{2 f_\pi}$, where $m_V = 800\,\mathrm{MeV}$ is an averaged vector meson mass and $f_\pi = 93~\mathrm{MeV}$ is the pion decay constant. The vector meson matrix $V$ is defined as 
\begin{equation}
\label{eq:vector meson}
   V = \begin{pmatrix}
            \frac{\rho^0}{\sqrt{2}} + \frac{\omega}{\sqrt{2}}  & \rho^+ & K^{* +}  \\
            \rho^- & -\frac{\rho^0}{\sqrt{2}} + \frac{\omega}{\sqrt{2}}   & K^{* 0} \\
           K^{* -} & \bar{K}^{* 0}  & \phi  
         \end{pmatrix}.
\end{equation}
In the limit where the three momenta of the external mesons are small compared to the vector mass, $V^\nu$ in Eq.~\eqref{eq:VVV} must correspond to the exchanged vector. Indeed, if it were an external vector, the index $\nu$ would necessarily be spatial, since $\epsilon^0$ is $0$ in the limit of $p\to 0$. In that case, $\partial_{\nu}$ would produce a three-vector, which vanishes in this limit. Therefore, $V^\nu$ cannot represent an external vector field and must instead correspond to the exchanged vector. It will then be contracted with $V^{\nu'}$ from the lower vertex in the diagram of Fig.~\ref{fig:MB} (see Eq.~\eqref{eq:VBB}) and the propagator yields
\begin{equation}
\label{eq:internal_prop}
\wick{ \c1V^{\nu} \c1V^{\nu'}}\to(-g^{\nu\nu'}+\frac{q^{\nu}q^{\nu'}}{M_V^2})\frac{i}{q^2-M_V^2+i\epsilon} \, .
\end{equation}
Near threshold, where we are working, $q^\nu, q^{\nu'} \to 0$, so the propagator reduces effectively to $-g^{\nu\nu'}$. This tensor is then contracted with $\partial_\nu$ at the upper vertex and with $\gamma_{\nu'}$ at the lower vertex, as we shall see below. Since $\partial_{i}$ ($i=1,2,3$) will be zero, only the $\nu=0$ component survives. Furthermore, for the external vectors we have $V^{\mu}V_{\mu}\to \epsilon^{\mu}\epsilon'_{\mu}\equiv -\vec \epsilon \cdot \vec \epsilon\,'$ since $\epsilon^0=0$ for external vectors. We estimate that the errors induced by approximating the external three-momenta as zero in the bound-state calculation are of order $(p/m_{\bar K^*})^2$, where $p$ is an averaged momentum of the constituent particles inside the molecule ($p \sim 200$~MeV). This amounts to roughly $5\%$. In practice, this correction is expected to be even smaller, as shown in the Appendix of Ref.~\cite{Sakai:2017hpg}.

For the lower vertex, the $VBB$ one, we calculate it using an analogous approach to Refs.~\cite{Debastiani:2017ewu,Wang:2022aga,Roca:2024nsi}. The vertex is given in terms of quarks and sandwiched between the baryon quark wave functions. The vertex is written as
\begin{align}
    \tilde{\mathcal{L}}_{\mathrm{VBB}} &= g q \bar{q}\gamma^{\nu'}V_{\nu'}, \label{eq:VBB}
\end{align}
where $q\bar{q}$ represents the quark wave function of the exchanged vector meson. At low energies, it follows from Eq.~\eqref{eq:internal_prop} that $\nu'=0$. This is further supported by the fact that, in this energy regime, $\gamma^{\nu'}$ is dominated by its time component, $\gamma^0$, since $\gamma^0 \sim 1$, while the spatial components are suppressed. Then we obtain the corresponding potential
\begin{align}
  V_{ij} &= \frac{1}{4f_\pi^2}C_{ij}(k^0+k'^0)\vec{\epsilon}~\vec{\epsilon}~', \label{eq:V}
\end{align}
where $k^0$, $k'^0$ ($\vec{\epsilon}$, $\vec{\epsilon}~'$) are the energies in the meson-baryon rest frame (spin
polarization vectors) of the initial and final mesons, respectively. The use of Eq.~\eqref{eq:VBB} to evaluate the lower $VBB$ vertex is relatively unconventional. More frequently, this vertex has been treated using effective Lagrangians. For baryons belonging to the octet of $1/2^+$ states, the corresponding expression is relatively straightforward~\cite{Klingl:1997kf,Palomar:2002hk,Oset:2010tof}\footnote{In Refs.~\cite{Palomar:2002hk,Oset:2010tof}, a misprint in the formalism of Ref.~\cite{Klingl:1997kf} is corrected.}. However, when the baryons in the $VBB$ vertex belong to the $3/2^+$ baryon decuplet, the formalism becomes considerably more cumbersome~\cite{Jenkins:1991es,Kolomeitsev:2003kt,Sarkar:2004jh}. The equivalence of Eq.~\eqref{eq:VBB} with the conventional $VBB$ vertex for octet baryons has been demonstrated explicitly in Refs.~\cite{Debastiani:2017ewu,Roca:2024nsi}. In addition, by comparing the results obtained using Eq.~\eqref{eq:VBB} with those of Ref.~\cite{Sarkar:2010saz}, we have verified that this equivalence also holds for the case of decuplet baryons.

The kernel of Eq.~\eqref{eq:V} corresponds to an $s$-wave and spin-independent potential (i.e., $\gamma^{\nu'} \to \gamma^{0} \sim 1$). Hence, the allowed spin–parity assignments are $J^P = 1/2^-$, $3/2^-$, and $5/2^-$. In this case, all spin-parity channels receive identical contributions, leading to a degenerate spectrum in these $J^P$ sectors. The coefficient matrix $C_{ij}$ with the channels $\bar{K}^*\Xi^*~(I=0)$ (1), $\omega\Omega$ (2) and $\phi\Omega$ (3), is given by
\begin{equation}
\label{eq:C}
   C_{ij} = \begin{pmatrix}
            0  & \sqrt{3} & -\sqrt{6}  \\
            ~ & 0   & 0 \\
           ~ & ~  & 0  
         \end{pmatrix},
\end{equation}
which is the same as Table 8 of Ref.~\cite{Sarkar:2010saz}, but a global sign difference because of the phase factor introduced by coupling here $VB$, instead of $BV$ taken in Ref.~\cite{Sarkar:2010saz}. The isospin $I=0$ $\bar{K}^*\Xi^*$ is given with the isospin phase convention $(\bar{K}^{*0},~-{K}^{*-})$ and $(\Xi^{*0},~\Xi^{*-})$ as
\begin{align}
  \lvert \bar{K}^*\Xi^*,~I=0\rangle= \frac{1}{\sqrt{2}}(\bar{K}^{*0}\Xi^{*-}+K^{*-}\Xi^{*0}). \label{eq:KXipc}
\end{align}
The scattering amplitude is then obtained by solving the Bethe–Salpeter equation
\begin{align}
  T &= [1-VG]^{-1}V, \label{eq:BSE}
\end{align}
where $G$ is the loop function 
\begin{align}
  G_{i}(\sqrt{s}) &= 2M_i\int_{|\vec{q}~|<q_{\textrm{max}}}\frac{d^3q}{(2\pi)^3}\frac{\omega_{i}(\vec{q}~)+E_{i}(\vec{q}~)}{2\omega_{i}(\vec{q}~)~E_{i}(\vec{q}~)}\frac{1}{s-(\omega_{i}(\vec{q}~)+E_{i}(\vec{q}~))^2+i\epsilon}, \label{eq:G}
\end{align}
with $M_i$ the baryon mass and $\omega_{i}(\vec{q}~)$, $E_{i}(\vec{q}~)$ the meson, baryon energy, respectively. For the channel $\bar{K}^*\Xi^*$, we also consider the mass distributions of $\bar{K}^*$ and $\Xi^*$, and perform a double convolution
\begin{align}
  \tilde{G}_{\bar{K}^*\Xi^*}(\sqrt{s}) =& \frac{1}{N}\int^{(m_{\bar{K}^*}+2\Gamma_{\bar{K}^*})^2}_{(m_{\bar{K}^*}-2\Gamma_{\bar{K}^*})^2} d\tilde{m}^2 \big(-\frac{1}{\pi}\big) \textrm{Im}\big(\frac{1}{\tilde{m}^2-m_{\bar{K}^*}^2+im_{\bar{K}^*}\Gamma_{\bar{K}^*}}\big) \notag
  \\&\times\int^{M_{\Xi^*}+2\Gamma_{\Xi^*}}_{M_{\Xi^*}-2\Gamma_{\Xi^*}} d\tilde{M} \big(-\frac{1}{\pi}\big) \textrm{Im}\big(\frac{1}{\tilde{M}-M_{\Xi^*}+i\frac{\Gamma_{\Xi^*}}{2}}\big)G_{\bar{K}^*\Xi^*}(\tilde{m},\tilde{M},\sqrt{s}), \label{eq:G1}
\end{align}
with 
\begin{align}
  N=&\int^{(m_{\bar{K}^*}+2\Gamma_{\bar{K}^*})^2}_{(m_{\bar{K}^*}-2\Gamma_{\bar{K}^*})^2} d\tilde{m}^2 \big(-\frac{1}{\pi}\big) \textrm{Im}\big(\frac{1}{\tilde{m}^2-m_{\bar{K}^*}^2+im_{\bar{K}^*}\Gamma_{\bar{K}^*}}\big) \notag
  \\&\times\int^{M_{\Xi^*}+2\Gamma_{\Xi^*}}_{M_{\Xi^*}-2\Gamma_{\Xi^*}} d\tilde{M} \big(-\frac{1}{\pi}\big) \textrm{Im}\big(\frac{1}{\tilde{M}-M_{\Xi^*}+i\frac{\Gamma_{\Xi^*}}{2}}\big). \label{eq:N}
\end{align}
Thus, the $G$ matrix entering in Eq.~\eqref{eq:BSE} can be expressed as
\begin{equation}
\label{eq:GM}
   G = \begin{pmatrix}
            \tilde{G}_{\bar{K}^*\Xi^*}  & ~ & ~  \\
            ~ & G_{\omega\Omega}   & ~ \\
           ~ & ~  & G_{\phi\Omega}  
         \end{pmatrix}.
\end{equation}

\subsection{Box Diagrams}
\begin{figure}[h!]  
    \centering
    \includegraphics[width=1\textwidth]{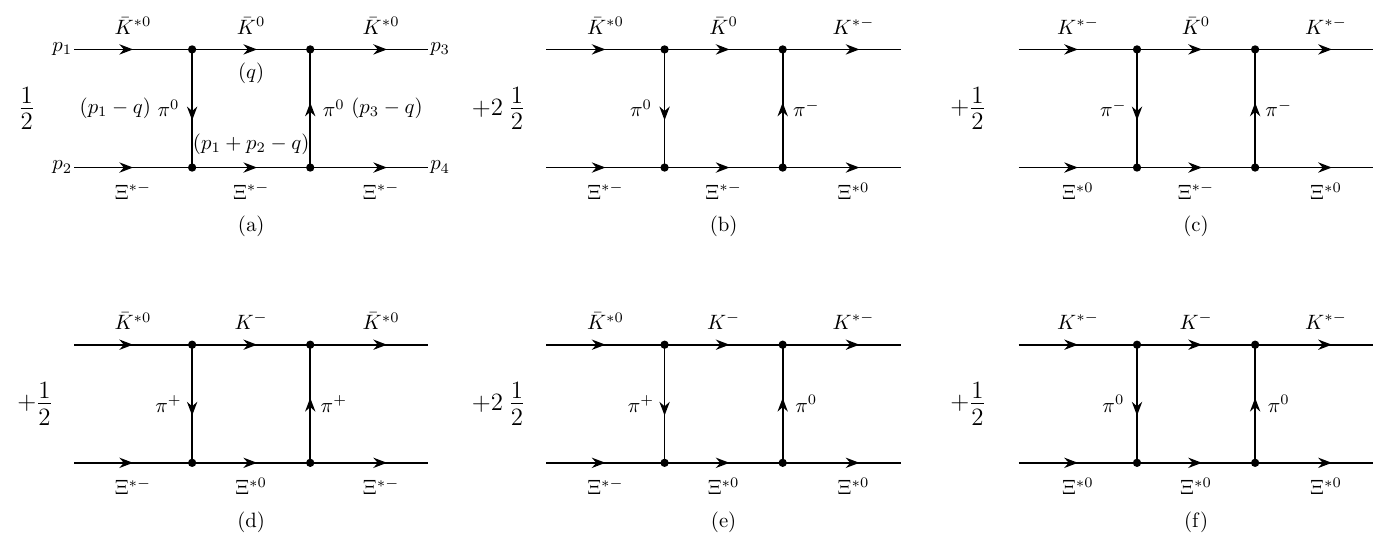} 
    \caption{Box diagrams for $\bar{K}^*\Xi^*$ decay into $\bar{K}\Xi^*$.}
    \label{fig:KXistar}
\end{figure}

\begin{figure}[h!]  
    \centering
    \includegraphics[width=1\textwidth]{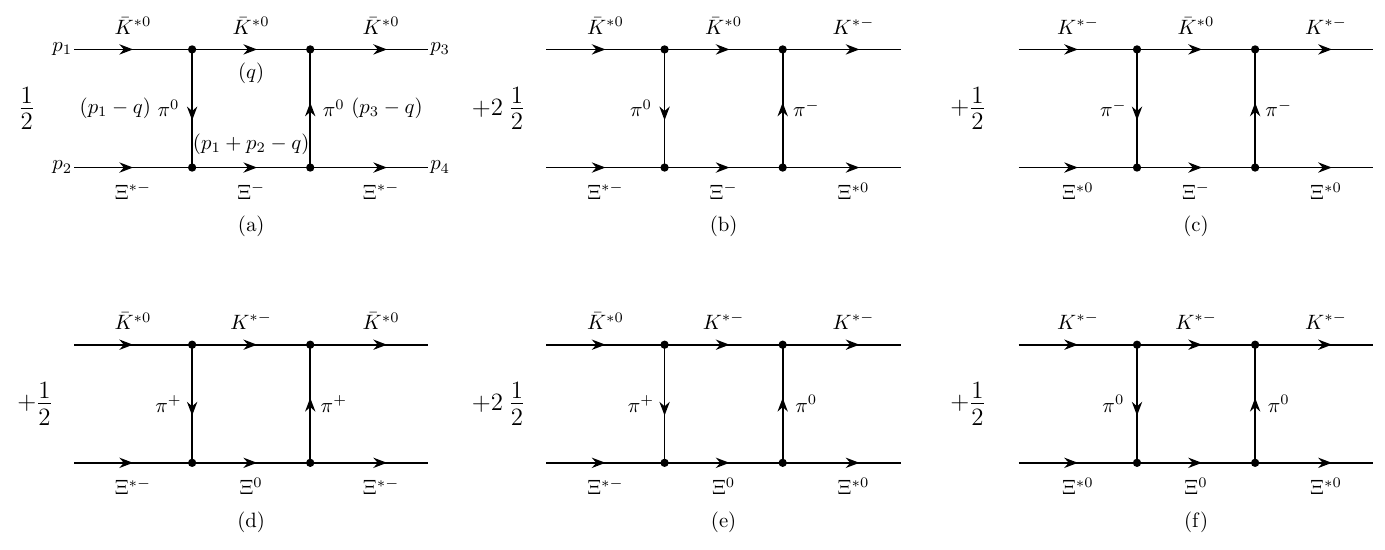} 
    \caption{Box diagrams for $\bar{K}^*\Xi^*$ decay into $\bar{K}^*\Xi$.}
    \label{fig:KstarXi}
\end{figure}

The inclusion of the $\bar{K}^*$ and $\Xi^*$ widths in Eq.~\eqref{eq:G1} already provides a finite width to the possible bound states generated in the coupled-channel approach. However, in order to establish a connection with the experimentally observed $\Omega(2380)$, it is necessary to account explicitly for its dominant decay modes. According to the PDG~\cite{ParticleDataGroup:2024cfk}, these are $\Xi^- \pi^+ K^-$, $\Xi(1530)^0 K^-$, and $\Xi^- \bar{K}^{*0}$. Within the present framework, the two-body decay channels arise naturally through the transitions $\bar{K}^* \Xi^* \to \bar{K} \Xi^*$ and $\bar{K}^* \Xi^* \to \bar{K}^* \Xi$. Their contribution is incorporated by evaluating the corresponding box diagrams, with $\bar{K} \Xi^*$ and $\bar{K}^* \Xi$ as intermediate states. The use of box diagrams to account for the width of dynamically generated states is well established in the literature~\cite{Molina:2008jw,Geng:2008gx,Molina:2020hde,Dai:2021vgf,Dai:2022ulk,Ikeno:2021mcb}. In this work, only the imaginary part of the box diagrams is retained, since the real part has been shown to be small and can be neglected~\cite{Molina:2008jw,Geng:2008gx}.
The box diagrams corresponding to the $\bar{K}^* \Xi^*$ decays into $\bar{K} \Xi^*$ and $\bar{K}^* \Xi$ are shown in Figs.~\ref{fig:KXistar} and~\ref{fig:KstarXi}, respectively.

\subsubsection{Box diagrams with $\bar{K}\Xi^*$ intermediate states}\label{Sec:KXistar}
In total, six box diagrams contribute to this process. However, it is sufficient to evaluate a single diagram explicitly, since the remaining ones can be obtained by applying isospin symmetry through the appropriate Clebsch--Gordan coefficients. 

Furthermore, since we are working close to threshold, the three-momenta of the external particles can be neglected in comparison with the vector-meson mass. We therefore set
\begin{equation}
\vec{p}_1 = \vec{p}_2 = \vec{p}_3 = \vec{p}_4 = 0 \, .
\end{equation}

The evaluation of the box diagrams in Fig.~\ref{fig:KXistar} requires two types of vertices: the ordinary $VPP$ coupling and the $\pi\Xi^*\Xi^*$ coupling. We first focus on the diagram of Fig.~\ref{fig:KXistar}(a), which contains four vertices $\bar{K}^{*0}\to \pi^0\bar{K}^0$, $\pi^0\bar{K}^0 \to\bar{K}^{*0} $, $\Xi^{*-}\to\pi^0\Xi^{*-}$, $\pi^0\Xi^{*-}\to\Xi^{*-}$.
The upper two $VPP$ vertices are obtained from the Lagrangian
\begin{align}
    \mathcal{L}_{\mathrm{VPP}} &= -i g \left\langle \left[P,\partial_\mu P\right]~ V^{\mu} \right\rangle, \label{eq:VPP}
\end{align}
with the matrix of the pseudoscalar mesons given by
\begin{equation}
\label{eq:p meson}
   P = \begin{pmatrix}
            \frac{\pi^0}{\sqrt{2}} + \frac{\eta}{\sqrt{3}}  & \pi^+ & K^{ +}  \\
            \pi^- & -\frac{\pi^0}{\sqrt{2}} + \frac{\eta}{\sqrt{3}}   & K^{ 0} \\
           K^{ -} & \bar{K}^{0}  & -\frac{\eta}{\sqrt{3}}  
         \end{pmatrix},
\end{equation}
where the standard $\eta$--$\eta^{\prime}$ mixing is taken into account, following Ref.~\cite{Bramon:1992kr}, and the $\eta^{\prime}$ contribution is subsequently neglected, since it is too massive to play a role in the present analysis. Hence the $\bar{K}^{*0}\to \pi^0\bar{K}^0$ vertex goes as
\begin{align}
    -it=i \mathcal{L}_{\mathrm{VPP}} =i\sqrt{2}~g~\vec{\epsilon}_{\bar{K}^{*0}}\vec{q},
    \label{eq:VPP1}
\end{align}
where we take $\epsilon^0_{\bar{K}^{*0}}=0$ since the $\bar{K}^*$ three momentum is neglected. The $\pi^0\bar{K}^0 \to\bar{K}^{*0} $ vertex is described by the same amplitude.

Next, we require the $\pi \Xi^* \Xi^*$ vertex. To this end, we begin by considering the coupling of $\pi$ with two $\frac{1}{2}^+$ baryons, assuming the $\pi$ momentum $\vec{q}$ in the $z$ direction and entering the baryon line, which is given by~\cite{Yang:2024nss}   
\begin{align}
    -it_{\pi BB'} &= \vec{\sigma}\vec{q} \left\langle B'\left|\tilde{V}_\pi\right|B \right\rangle, \label{eq:PBB}
\end{align}
where $B$, $B'$ denote the quark wave functions of baryons. The operator $\tilde{V}_\pi$ reads
\begin{align}
    \tilde{V}_\pi =\frac{3}{5}\frac{f_{\pi NN}}{m_\pi}\sqrt{2}\sum_i\left\{ \begin{aligned}
 &u\bar{d} \\
\frac{1}{\sqrt{2}} (&u\bar{u}-d\bar{d})\\
 &d\bar{u}
\end{aligned}
 \right\}_i\sigma_{z,i}~, \label{eq:Vpi}
\end{align}
where the index $i$ runs over the quarks in the baryon, $f_{\pi NN}$ is the $\pi^0pp$ coupling constant (set to $f_{\pi NN}=1$ in the present work~\cite{Ericson:1988gk}), and $m_\pi$ is the pion mass. In Eq.~\eqref{eq:Vpi}, $\sigma_z$ is the ordinary quark spin matrix for spin $\frac{1}{2}$ in $z$ direction. The three components in the bracket of Eq.~\eqref{eq:Vpi} stand for $\pi^+$, $\pi^0$, $\pi^-$ fields, respectively. This operator for the case of the $\pi^0 pp$ vertex, leads to the macroscopic vertex
\begin{align}
    -it &= \frac{f_{\pi NN}}{m_\pi}\vec{\sigma}\vec{q} , \label{eq:pipp}
\end{align}
where $\vec{\sigma}$ denotes the spin operator of the proton.

We now turn to the $\pi \Xi^* \Xi^*$ coupling. At the macroscopic level for $J^P = \frac{3}{2}^+$ baryons, the spin operator is $\tilde{S} \cdot \vec{q}$. In the spherical basis, this operator can be written as
\begin{align}
    \tilde{S}\cdot\vec{q} =\sum_\mu(-1)^\mu\tilde{S}_\mu q_{-\mu}, \label{eq:sq}
\end{align}
with
$
\begin{cases}
q_{+1}&=-\frac{1}{\sqrt{2}}(q_x+iq_y) \\
q_{-1}&=\frac{1}{\sqrt{2}}(q_x-iq_y)\\
q_0&=q_z
\end{cases}$
and
$
\begin{cases}
\tilde{S}_{+1}&=-\frac{1}{\sqrt{2}}({S}_x+i{S}_y) \\
\tilde{S}_{-1}&=\frac{1}{\sqrt{2}}({S}_x-i{S}_y)\\
\tilde{S}_0&={S}_z
\end{cases}$.

To evaluate the $\Xi^{*-}\to\pi^0\Xi^{*-}$ vertex shown in Fig~\ref{fig:KXistar}(a), Eq.~\eqref{eq:Vpi} is used together with the spin-flavor quark wave function
\begin{align}
    \Phi_{\Xi^{*-}} &= \frac{1}{\sqrt{3}}\left| dss+sds+ssd \right>\chi_s, \label{eq:phiXi}
\end{align}
with $\chi_s$ the spin symmetric wave function.

The matrix elements of the spin transition operator $\tilde{S}_\mu$ are expressed in terms of Clebsch-Gordan coefficients using the Wigner-Eckart theorem, 
\begin{align}
   \left<M'\left|\tilde{S}_\mu\right|M\right>=\alpha~\mathcal{C}(\frac{3}{2}~1~\frac{3}{2};M~\mu~M'), \label{eq:St}
\end{align}
which properly evaluated with $\mu=0$ and $M=M'=\frac{3}{2}$ renders
\begin{align}
   \alpha=\frac{1}{2}\sqrt{15}~. \label{eq:alpha value}
\end{align}

Next, taking into account Eqs.~\eqref{eq:Vpi}, \eqref{eq:phiXi} and the $\chi_s$ wave function of \cite{Close1979}, one finds
\begin{align}
   \left<\Xi^{*-},\tilde{S}_z=\frac{3}{2}\left|\tilde{V}_\pi\right|\Xi^{*-},\tilde{S}_z=\frac{3}{2}\right>=-\frac{3}{5}\frac{f_{\pi NN}}{m_\pi}. \label{eq:XiVpi}
\end{align}
This result can be matched to a macroscopic operator of the form $\left<M'\left|~\beta\tilde{S}\vec{q}~\right|M\right>$, which fixes $\beta=-\frac{2}{5}\frac{f_{\pi NN}}{m_\pi}$. Accordingly, we obtain
\begin{align}
    -it=-\frac{2}{5}\frac{f_{\pi NN}}{m_\pi}\tilde{S}\cdot\vec{q}. \label{eq:tXiXi}
\end{align}
The $\Xi^{*-}\pi^0\to\Xi^{*-}$ vertex has the same coupling strength, with an additional minus sign arising from the reversed momentum flow of the $\pi^0$. Combining both vertices, the contribution of the two lower vertices is given by 
\begin{align}
    -\frac{4}{25}(\frac{f_{\pi NN}}{m_\pi})^2\sum_M\left<S_4\left|\tilde{S}\cdot\vec{q}~\right|M\right>\left<M\left|\tilde{S}\cdot\vec{q}~\right|S_2\right>, \label{eq:lower}
\end{align}
where $S_2$, $S_4$ and $M$ are the third component of the spin of the initial, final and intermediate $\Xi^{*-}$, respectively. 

The diagrams of Fig.~\ref{fig:KXistar} are evaluated by averaging over spins for diagonal transitions, $S_2=S_4$. In general, one has
\begin{align}
    \left<S_4\left|\tilde{S}\cdot\vec{q}~\right|M\right>&\left<M\left|\tilde{S}\cdot\vec{q}~\right|S_4\right>=\left<S_4\left| \sum_\mu (-1)^\mu S_\mu q_{-\mu} \right|M\right>\left<M\left|\sum_\nu (-1)^\nu S_{-\nu} q_{\nu}\right|S_4\right>\notag\\
    &=\sum_\mu (-1)^\mu ~\alpha ~\mathcal{C}(\frac{3}{2}~1~\frac{3}{2};M~\mu~S_4)q_{-\mu} \sum_\nu (-1)^\nu ~\alpha~ \mathcal{C}(\frac{3}{2}~1~\frac{3}{2};S_4~-\nu~M)q_\nu,
    \label{eq:StCG2}
\end{align}
where conditions $M+\mu=S_4$ and $S_4-\nu=M$ enforce $\mu=\nu$. Permuting the indices $S_4$, $M$ in the second Clebsch-Gordan coefficient~\cite{Rose1957}, one obtains
\begin{align}
&\sum_\mu(-1)^\mu~\alpha^2~q_{-\mu}~q_{\mu}~\mathcal{C}(\frac{3}{2}~1~\frac{3}{2};M~\mu~S_4)~\mathcal{C}(\frac{3}{2}~1~\frac{3}{2};M~\mu~S_4).
    \label{eq:StCG3}
\end{align}

Summing Eq.~\eqref{eq:StCG3} over the intermediate spin projection $M$, the orthogonality of the Clebsch–Gordan coefficients yields $\delta_{\frac{3}{2}\frac{3}{2}}$ and one finally finds
\begin{align}
    \frac{1}{4}\sum_{S_4}\sum_M\left<S_4\left|\tilde{S}\cdot\vec{q}~\right|M\right>\left<M\left|\tilde{S}\cdot\vec{q}~\right|S_4\right>=\frac{1}{4}~\alpha^2~\sum_\mu(-1)^\mu q_{-\mu}~q_\mu=\frac{1}{4}~\alpha^2~\vec{q}~^2=\frac{15}{16}~\vec{q}~^2.
    \label{eq:StCG4}
\end{align}
Substituting this result into Eq.~\eqref{eq:lower}, we obtain
\begin{equation}
    \begin{aligned}
        &(1) \Xi^{*-}\overset{\pi}\to\Xi^{*-}\overset{\pi}\to\Xi^{*-}: -\frac{4}{25}(\frac{f_{\pi NN}}{m_\pi})^2\frac{15}{16}~\vec{q}~^2;\\
        &(2) \Xi^{*-}\overset{\pi}\to\Xi^{*-}\overset{\pi}\to\Xi^{*0}: \frac{4\sqrt{2}}{25}(\frac{f_{\pi NN}}{m_\pi})^2\frac{15}{16}~\vec{q}~^2;\\
&(3) \Xi^{*0}\overset{\pi}\to\Xi^{*-}\overset{\pi}\to\Xi^{*0}: -\frac{8}{25}(\frac{f_{\pi NN}}{m_\pi})^2\frac{15}{16}~\vec{q}~^2;\\
&(4) \Xi^{*-}\overset{\pi}\to\Xi^{*0}\overset{\pi}\to\Xi^{*-}: -\frac{8}{25}(\frac{f_{\pi NN}}{m_\pi})^2\frac{15}{16}~\vec{q}~^2;\\
&(5) \Xi^{*-}\overset{\pi}\to\Xi^{*0}\overset{\pi}\to\Xi^{*0}: -\frac{4\sqrt{2}}{25}(\frac{f_{\pi NN}}{m_\pi})^2\frac{15}{16}~\vec{q}~^2;\\
&(6) \Xi^{*0}\overset{\pi}\to\Xi^{*0}\overset{\pi}\to\Xi^{*0}: -\frac{4}{25}(\frac{f_{\pi NN}}{m_\pi})^2\frac{15}{16}~\vec{q}~^2.
    \end{aligned}
    \label{eq:Amp2}
\end{equation}

We next turn to the upper vertices of the diagrams, which are of the type given in Eq.~\eqref{eq:VPP1}. The resulting structure of box diagrams is of the form
\begin{equation}
\int d^3q\vec{\epsilon}~\vec{q}~\vec{\epsilon}~'\vec{q}f(q^2)=\int d^3q\epsilon_i\epsilon'_jq_iq_jf(q^2)=\frac{1}{3}\epsilon_i\epsilon'_j\delta_{ij}\int d^3q~ \vec{q}~^2f(q^2) \, ,
\end{equation}
 which allows one to write the contribution of the two upper vertices as
\begin{equation}
    \begin{aligned}
        &(1) \bar{K}^{*0}\overset{\pi}\to\bar{K}^0\overset{\pi}\to\bar{K}^{*0}: -\frac{2}{3}g^2\vec{q}~^2\vec{\epsilon}~\vec{\epsilon}~';\\
        &(2) \bar{K}^{*0}\overset{\pi}\to\bar{K}^0 \overset{\pi}\to{K}^{*-}: \frac{2\sqrt{2}}{3}g^2\vec{q}~^2\vec{\epsilon}~\vec{\epsilon}~';\\
        &(3) {K}^{*-}\overset{\pi}\to\bar{K}^0 \overset{\pi}\to{K}^{*-}: -\frac{4}{3}g^2\vec{q}~^2\vec{\epsilon}~\vec{\epsilon}~';\\
        &(4) \bar{K}^{*0}\overset{\pi}\to{K}^- \overset{\pi}\to\bar{K}^{*0}: -\frac{4}{3}g^2\vec{q}~^2\vec{\epsilon}~\vec{\epsilon}~';\\
&(5) \bar{K}^{*0}\overset{\pi}\to{K}^- \overset{\pi}\to{K}^{*-}: -\frac{2\sqrt{2}}{3}g^2\vec{q}~^2\vec{\epsilon}~\vec{\epsilon}~';\\
&(6) {K}^{*-}\overset{\pi}\to{K}^- \overset{\pi}\to{K}^{*-}: -\frac{2}{3}g^2\vec{q}~^2\vec{\epsilon}~\vec{\epsilon}~'.
    \end{aligned}
    \label{eq:Amp1}
\end{equation}

Having evaluated all the vertices, we now compute the total amplitude of the box diagram in Fig.~\ref{fig:KXistar}. Starting with Fig.~\ref{fig:KXistar}(a), and setting $\vec{p}_1=\vec{p}_2=\vec{p}_3=\vec{p}_4=0$, the corresponding amplitude reads
\begin{align}
    \tilde{V}_{\bar{K}\Xi^*(a)} =& i\int\frac{d^4q}{(2\pi)^4} (-\frac{2}{3}g^2~\vec{q}~^2)[-\frac{4}{25}(\frac{f_{\pi NN}}{m_\pi})^2\frac{15}{16}~\vec{q}~^2]
    \frac{i}{(p_1-q)^2-m_{\pi}^2+i\epsilon}\frac{i}{(p_3-q)^2-m_{\pi}^2+i\epsilon} \notag \\ &\times\frac{1}{2\omega_{\bar{K}}(\vec{q}~)}\frac{i}{q^0-\omega_{\bar{K}}(\vec{q}~)+i\epsilon}~\frac{M_{\Xi^*}}{E_{\Xi^*}(\vec{q}~)} \frac{i}{p_1^0+p_2^0-q^0-E_{\Xi^*}(\vec{q}~)+i\epsilon}\vec{\epsilon}~\vec{\epsilon}~' , \label{eq:VK1}
\end{align}
with 
\begin{align}
    p_1^0=p_3^0=&\frac{M_{inv}^2+m^2_{\bar{K}^*}-M^2_{\Xi^*}}{2M_{inv}}, \label{eq:p1}
\\
    p_2^0=p_4^0=&\frac{M_{inv}^2+M^2_{\Xi^*}-m^2_{\bar{K}^*}}{2M_{inv}}. \label{eq:p2}
\end{align}

As discussed above, we are only interested in the imaginary part of the diagram that arises when the intermediate $\bar{K}\Xi^*$ states are placed on shell. For this purpose, it is convenient to separate the positive- and negative-energy parts of the propagator as
\begin{align}
    \frac{1}{q^2-m^2+i\epsilon}=\frac{1}{2\omega(\vec{q}~)}\left(\frac{1}{q^0-\omega(\vec{q}~)+i\epsilon}-\frac{1}{q^0+\omega(\vec{q}~)-i\epsilon}\right) \label{eq:p1}
\end{align}
and retain only the first term. This prescription has already been implemented in Eq.~\eqref{eq:VK1}. For the baryon propagator we have the same, but keeping the field normalization of Mandl and Shaw~\cite{MandlShaw2010}. For the exchanged pions, which are necessarily off shell, we keep the full propagators. Then, we can perform an analytical integration over the variable $q^0$, using Cauchy’s residue theorem (alternatively, one may use Cutkosky rules). As a result, Eq.~\eqref{eq:VK1} takes the form
\begin{align}
    \tilde{V}_{\bar{K}\Xi^*(a)} =& \int\frac{d^3q}{(2\pi)^3} (-\frac{2}{3}g^2~\vec{q}~^2)[-\frac{4}{25}(\frac{f_{\pi NN}}{m_\pi})^2\frac{15}{16}~\vec{q}~^2]
    \left(\frac{1}{(p_1^0-\omega_{\bar{K}}(\vec{q}~))^2-\omega_{\pi}(\vec{q}~)^2+i\epsilon}\right)^2 \notag \\ &\times\frac{1}{2\omega_{\bar{K}}(\vec{q}~)}~\frac{M_{\Xi^*}}{E_{\Xi^*}(\vec{q}~)} \frac{1}{p_1^0+p_2^0-\omega_{\bar{K}}(\vec{q}~)-E_{\Xi^*}(\vec{q}~)+i\epsilon}\vec{\epsilon}~\vec{\epsilon}~'. \label{eq:VK12}
\end{align}
Using $i~\textrm{Im}(\frac{1}{x+i\epsilon})=-i\pi\delta(x)$ we then obtain
\begin{align}
    \textrm{Im}\tilde{V}_{\bar{K}\Xi^*(a)} =& -\frac{M_{\Xi^*}}{4\pi}\frac{\vec{q}}{M_{inv}} (-\frac{2}{3}g^2~\vec{q}~^2)[-\frac{4}{25}(\frac{f_{\pi NN}}{m_\pi})^2\frac{15}{16}~\vec{q}~^2]
    \left(\frac{1}{(p_1^0-\omega_{\bar{K}}(\vec{q}~))^2-\omega_{\pi}(\vec{q}~)^2+i\epsilon}\right)^2\vec{\epsilon}~\vec{\epsilon}~' \notag\\
    =&-\frac{1}{10}\vec{\epsilon}~\vec{\epsilon}~'g^2~\vec{q}~^5\frac{M_{\Xi^*}}{4\pi}(\frac{f_{\pi NN}}{m_\pi})^2\frac{1}{M_{inv}}\left(\frac{1}{(p_1^0-\omega_{\bar{K}}(\vec{q}~))^2-\omega_{\pi}(\vec{q}~)^2+i\epsilon}\right)^2
    , \label{eq:imVK1}
\end{align}
with
\begin{equation}
    |\vec{q}~| = \frac{\lambda^{\frac{1}{2}}\left(M_{inv}^2~, m^2_{\bar{K}}~, M^2_{\Xi^*}\right)}{2M_{inv}}. \label{eq:qv1}
\end{equation}
Summing the six diagrams shown in Fig.~\ref{fig:KXistar}, with the corresponding weights of Eqs.~\eqref{eq:Amp2} and \eqref{eq:Amp1}, the total imaginary part of the amplitude is written as
\begin{align}
    \textrm{Im}\tilde{V}_{\bar{K}\Xi^*} 
    =&-\frac{9}{10}\vec{\epsilon}~\vec{\epsilon}~'g^2~\vec{q}~^5\frac{M_{\Xi^*}}{4\pi}(\frac{f_{\pi NN}}{m_\pi})^2\frac{1}{M_{inv}}\left(\frac{1}{(p_1^0-\omega_{\bar{K}}(\vec{q}~))^2-\omega_{\pi}(\vec{q}~)^2+i\epsilon}\right)^2FF(\vec{q}~)^2 \, ,\label{eq:imVK1}
\end{align}
where, since the pions are off shell, we introduce a form factor $FF(\vec{q}~)$, defined as~\cite{Dai:2022ulk}
\begin{align}
    FF(\vec{q}~)=& e^{[(p^0_1-q^0)^2-\vec{q}~^2]/\Lambda^2}, \label{eq:FF}
\end{align}
with 
\begin{align}
    q^0=\frac{M_{inv}^2+m_{\bar{K}}^2-M^2_{\Xi^*}}{2M_{inv}} \,. \label{eq:q01}
\end{align} 
In the numerical evaluation we take $\Lambda\simeq 1~\text{GeV}$. A form factor of this type, with a value of $\Lambda$ around $1$~GeV, has been tuned in previous works to obtain realistic widths from box diagrams in many cases~\cite{Geng:2008gx,Molina:2008jw,Molina:2020hde,Ikeno:2021mcb}.

\subsubsection{Box diagrams with $\bar{K}^*\Xi$ intermediate states}

To evaluate the box diagrams in Fig.~\ref{fig:KstarXi}, two additional types of vertices are required: the anomalous $VVP$ coupling and the $\pi\Xi\Xi^*$ coupling. The upper $VVP$ vertex is given by the Lagrangian~\cite{Bramon:1992kr}
\begin{align}
    \mathcal{L}_{\mathrm{VVP}} &= \frac{G'}{\sqrt{2}}\epsilon^{\mu \nu \alpha \beta} \left\langle \partial_\mu V_\nu \partial_\alpha V_{\beta} P \right\rangle, \label{eq:VVP}
\end{align}
with coupling constants $G'=\frac{3g'^2}{4\pi^2f_\pi}$, $g'=-\frac{G_vm_\rho}{\sqrt{2}f_\pi^2}$, $G_v=55$ MeV ($G'\approx14~\textrm{GeV}^{-1}$)~\cite{Bramon:1992kr,Oset:2002sh}.

We start with Fig.~\ref{fig:KstarXi}(a), the $\bar{K}^{*0}\to \pi^0\bar{K}^{*0}$ vertex goes as ($\epsilon^{\mu \nu\alpha \beta }=\epsilon^{\alpha \beta\mu \nu }$)
\begin{align}
    -it=i \mathcal{L}_{\mathrm{VVP}} =-i\frac{G'}{\sqrt{2}}\frac{1}{\sqrt{2}}\epsilon^{\alpha \beta\mu \nu }(-ip_{1\alpha}){\epsilon}_{\beta(in)}(iq_\mu){\epsilon}_{\nu(out)},
    \label{eq:VVP1}
\end{align}
where $in$ and $out$ represent the incoming (in this case the initial) and outgoing (in this case the intermediate) $\bar{K}^{*0}$. One should notice that since $\vec{p}_1=0$, we need to take $\alpha=0$, then $\epsilon^{\alpha \beta\mu \nu }=\epsilon^{0 \beta\mu \nu}=\epsilon^{ijk}$, and an extra minus sign comes from raising the covariant indices. So that Eq.~\eqref{eq:VVP1} becomes
\begin{align}
    -it=i\frac{G'}{\sqrt{2}}\frac{1}{\sqrt{2}}\epsilon^{ijk}~p_{1}^0~{\epsilon}_{(in)}^iq^j{\epsilon}_{(out)}^k.
    \label{eq:VVP2}
\end{align}
Similarly, the $\bar{K}^{*0} \pi^0\to\bar{K}^{*0}$ vertex is written with the same structure
\begin{align}
    -it=-i\frac{G'}{\sqrt{2}}\frac{1}{\sqrt{2}}\epsilon^{ijk}~p_{3}^0~{\epsilon}_{(in)}^iq^j{\epsilon}_{(out)}^k.
    \label{eq:VVP3}
\end{align}
But in this case $in$ and $out$ correspond to the intermediate and final $\bar{K}^{*0}$, respectively. When combining these two vertices via the intermediate $\bar{K}^{*0}$, we must contract the polarization vectors of this intermediate meson, that is ${\epsilon}_{(out)}^k$ in Eq.~\eqref{eq:VVP2} and ${\epsilon}_{(in)}^i$ in Eq.~\eqref{eq:VVP3}.

For the $\pi\Xi\Xi^*$ coupling, we introduce two spin transition operators at the hadron level: $\vec{S}\cdot\vec{q}$ for the transition from spin $\tfrac{3}{2}$ to $\tfrac{1}{2}$, and $\vec{S}^{+}\cdot\vec{q}$ for the inverse process. According to the Wigner--Eckart theorem, we have~\cite{Yang:2024nss}
\begin{align}
   \left\langle \tfrac{3}{2}\, M \left| S^{+}_{\mu} \right| \tfrac{1}{2}\, m \right\rangle
   = \mathcal{C}\!\left(\tfrac{1}{2}\, 1\, \tfrac{3}{2};\, m\, \mu\, M \right)
   \left\langle \tfrac{3}{2} \big\| \vec{S}^{+} \big\| \tfrac{1}{2} \right\rangle ,
   \label{eq:Sp}
\end{align}
where $\mu$ denotes the spherical component of $\vec{S}^{+}$. The reduced matrix element is set to unity:
$\left\langle \tfrac{3}{2} \big\| \vec{S}^{+} \big\| \tfrac{1}{2} \right\rangle \equiv 1$.

In Fig.~\ref{fig:KstarXi}(a), we evaluate the $\Xi^{*-}\to \pi^0\Xi^{-}$ and $\pi^0\Xi^{-}\to \Xi^{*-}$ vertices. Using Eq.~\eqref{eq:Vpi} with the quark wave function of the baryon, and mapping the quark-level operators to hadron spin operators as before, the vertex amplitudes are
\begin{align}
    -it_{\Xi^{*-}\to \pi^0\Xi^{-}}&=-\frac{2}{5}\sqrt{3}\frac{f_{\pi NN}}{m_\pi}\vec{S}\cdot(-\vec{q}~), \label{eq:tXi1}
\\
    -it_{\pi^0\Xi^{-}\to \Xi^{*-}}&=-\frac{2}{5}\sqrt{3}\frac{f_{\pi NN}}{m_\pi}\vec{S}^+\cdot\vec{q}. \label{eq:tXi2}
\end{align}

As in Sec.~\ref{Sec:KXistar}, we consider only diagonal transitions for $\Xi^*$ ($M=M'$) and average over the spin projections ($M$). Thus, we have
\begin{align}
   \frac{1}{4}\sum_M\sum_m\left<M\left|\vec{S}^+\vec{q}~\right|m\right>\left<m\left|\vec{S}~\vec{q}~\right|M\right> =\frac{1}{4}\sum_m\sum_M\left<m\left|\vec{S}~\vec{q}~\right|M\right>\left<M\left|\vec{S}^+\vec{q}~\right|m\right>
   \label{eq:sumchange}
\end{align}
and using the spin-sum identity,
\begin{align}
   \sum_M\left<m'\left|S_i\right|M\right> \left<M\left|S_j^+\right|m\right>=\left<m'\left|  \frac{2}{3}\delta_{ij}-\frac{i}{3}\epsilon_{ijk}\sigma_k \right|m\right>, \label{eq:SSp}
\end{align}
the spin sum for the two lower vertices gives an overall factor of $\frac{1}{4}2\frac{2}{3}\vec{q}~^2$. This leads to the contributions from the six possible lower-vertex transitions:
\begin{equation}
    \begin{aligned}
    &(1) \Xi^{*-}\overset{\pi}\to\Xi^{-}\overset{\pi}\to\Xi^{*-}: -\frac{4}{25}(\frac{f_{\pi NN}}{m_\pi})^2~\vec{q}~^2;\\
&(2) \Xi^{*-}\overset{\pi}\to\Xi^{-}\overset{\pi}\to\Xi^{*0}: \frac{4\sqrt{2}}{25}(\frac{f_{\pi NN}}{m_\pi})^2~\vec{q}~^2;\\
&(3) \Xi^{*0}\overset{\pi}\to\Xi^{-}\overset{\pi}\to\Xi^{*0}: -\frac{8}{25}(\frac{f_{\pi NN}}{m_\pi})^2~\vec{q}~^2;\\
&(4) \Xi^{*-}\overset{\pi}\to\Xi^{0}\overset{\pi}\to\Xi^{*-}: -\frac{8}{25}(\frac{f_{\pi NN}}{m_\pi})^2~\vec{q}~^2;\\
&(5) \Xi^{*-}\overset{\pi}\to\Xi^{0}\overset{\pi}\to\Xi^{*0}: -\frac{4\sqrt{2}}{25}(\frac{f_{\pi NN}}{m_\pi})^2~\vec{q}~^2;\\
&(6) \Xi^{*0}\overset{\pi}\to\Xi^{0}\overset{\pi}\to\Xi^{*0}: -\frac{4}{25}(\frac{f_{\pi NN}}{m_\pi})^2~\vec{q}~^2.
    \end{aligned}
    \label{eq:Amp4}
\end{equation}
Using these results, we now consider the upper VVP vertices in Eqs.~\eqref{eq:VVP2} and \eqref{eq:VVP3}. Summing over the polarization of the intermediate $\bar{K}^{*}$, the relevant contraction reads
\begin{equation}
    \begin{aligned}
    \sum_{\tilde{\epsilon}_{\textrm{pole}}}\epsilon^{ijk}\epsilon^{i'j'k'}\epsilon_iq_j\tilde{\epsilon}_k\epsilon'_{i'}q_{j'}\tilde{\epsilon}_{k'}&=\epsilon^{ijk}\epsilon^{i'j'k'}\epsilon_iq_j\epsilon'_{i'}q_{j'}\delta_{kk'}\\
&=\epsilon^{ijk}\epsilon^{i'j'k}\epsilon_iq_j\epsilon'_{i'}q_{j'}\\
&\to \epsilon^{ijk}\epsilon^{i'j'k}\epsilon_i\epsilon'_{i'}\frac{1}{3}\vec{q}~^2~\delta_{jj'}\\
&=\frac{2}{3}\vec{q}~^2\vec{\epsilon}~\vec{\epsilon}~',
    \end{aligned}
    \label{eq:epsilonsum}
\end{equation}
where, in the last step, we used the standard angular average $\int d^3q~q_iq_jf(q^2)=\frac{1}{3}\delta_{ij}\int d^3q~ \vec{q}~^2f(q^2)$, which is valid since the lower vertices now depend only on $\vec{q}~^2$. With this, the contribution from the upper vertices is given by
\begin{equation}
    \begin{aligned}
   &(1) \bar{K}^{*0}\overset{\pi}\to\bar{K}^{*0} \overset{\pi}\to\bar{K}^{*0}: -\frac{1}{3}(\frac{G'}{\sqrt{2}})^2(p_1^0)^2\vec{q}~^2~\vec{\epsilon}~\vec{\epsilon}~';\\
&(2) \bar{K}^{*0}\overset{\pi}\to\bar{K}^{*0} \overset{\pi}\to{K}^{*-}: \sqrt{2}\frac{1}{3}(\frac{G'}{\sqrt{2}})^2(p_1^0)^2\vec{q}~^2~\vec{\epsilon}~\vec{\epsilon}~';\\
&(3) {K}^{*-}\overset{\pi}\to\bar{K}^{*0} \overset{\pi}\to{K}^{*-}: -{2}\frac{1}{3}(\frac{G'}{\sqrt{2}})^2(p_1^0)^2\vec{q}~^2~\vec{\epsilon}~\vec{\epsilon}~';\\
&(4) \bar{K}^{*0}\overset{\pi}\to{K}^{*-} \overset{\pi}\to\bar{K}^{*0}: -{2}\frac{1}{3}(\frac{G'}{\sqrt{2}})^2(p_1^0)^2\vec{q}~^2~\vec{\epsilon}~\vec{\epsilon}~';\\
&(5) \bar{K}^{*0}\overset{\pi}\to{K}^{*-} \overset{\pi}\to{K}^{*-}: -\sqrt{2}\frac{1}{3}(\frac{G'}{\sqrt{2}})^2(p_1^0)^2\vec{q}~^2~\vec{\epsilon}~\vec{\epsilon}~';\\
&(6) {K}^{*-}\overset{\pi}\to{K}^{*-} \overset{\pi}\to{K}^{*-}: -\frac{1}{3}(\frac{G'}{\sqrt{2}})^2(p_1^0)^2\vec{q}~^2~\vec{\epsilon}~\vec{\epsilon}~',
    \end{aligned}
    \label{eq:Amp3}
\end{equation}
with $\vec{\epsilon}$ and $\vec{\epsilon}~'$ the spin polarization vectors of the initial and final $\bar{K}^{*}$, respectively.

Finally, summing all contributions from the six diagrams in Fig.~\ref{fig:KstarXi}, the imaginary part of the full $\bar{K}^*\Xi$ box amplitude is
\begin{align}
    \textrm{Im}\tilde{V}_{\bar{K}^*\Xi}=&-\frac{6}{25}\vec{\epsilon}~\vec{\epsilon}~'G'^2(p_1^0)^2~\vec{q}~^5\frac{M_{\Xi}}{4\pi}(\frac{f_{\pi NN}}{m_\pi})^2\frac{1}{M_{inv}}\left(\frac{1}{(p_1^0-\omega_{\bar{K}^*}(\vec{q}~))^2-\omega_{\pi}(\vec{q}~)^2+i\epsilon}\right)^2FF(\vec{q}~)^2
    , \label{eq:imVK1}
\end{align}
where the intermediate three-momentum and energy in the center-of-mass frame are
\begin{align}
    |\vec{q}~| &= \frac{\lambda^{\frac{1}{2}}\left(M_{inv}^2~, m^2_{\bar{K}^*}~, M^2_{\Xi}\right)}{2M_{inv}}, \label{eq:qv2}
\\
    q^0&=\frac{M_{inv}^2+m_{\bar{K}^*}^2-M^2_{\Xi}}{2M_{inv}}. \label{eq:q02}
\end{align}
These expressions are analogous to those of the $\bar{K}\Xi^*$ case (see Eqs.~\eqref{eq:qv1} and \eqref{eq:q01}), but differ due to the different intermediate masses.

At this point, we would like to make a clarification. The box diagrams contribute to all possible spins of the $\Omega^*$ state that we obtained, namely $1/2^-,3/2^-,5/2^-$. At first sight, one might think that the $\bar K \Xi^*$ intermediate states in the box diagrams can only generate $J^P = 3/2^-$, which would be the case if the $\bar K \Xi^*$ pair were in $s$-wave. However, this is not the case. The $\vec q$ dependence arising from the vertices of the box diagram in Eq.~\eqref{eq:VK1}, which behaves as $\vec q^{\,2}\vec q^{\,2}$, can be written as $q_i q_i$ for the upper vertices and $q_j q_j$ for the lower ones. The $\bar K \Xi^*$ propagation then involves a structure $q_i q_j$, which can be decomposed as $q_iq_j\equiv (q_iq_j-\frac{1}{3}\vec q^{\,2}\delta_{ij})+\frac{1}{3}\vec q^{\,2}\delta_{ij}$. This decomposition explicitly separates the $d$-wave component (the traceless tensor term) and the $s$-wave component (the $\delta_{ij}$ term) in the intermediate propagation. Consequently, the $\bar K \Xi^*$ intermediate states contain both $s$- and $d$-wave contributions, allowing all three spin states of the initial $\bar K^* \Xi^*$ system to be accommodated. The same reasoning applies to the box diagrams involving intermediate $\bar K^* \Xi$ states.

The box diagrams that we have calculated can break the degeneracy among the original spin states. This mechanism has been implemented in works such as~\cite{Garzon:2012np} in the study of the mixing between pseudoscalar-baryon and vector-baryon interactions. It has also been considered in the lifting of the spin degeneracy of the $\bar D^*\Sigma_c$ states, leading to the $P_c(4440)$ and $P_c(4457)$ in~\cite{Yang:2024nss}. In both cases, the breaking of the degeneracy results in relatively small shifts in the masses and widths, as can be seen from the difference between the two $P_c$ states mentioned above. It should be noted that in the present case we have not broken the degeneracy. This is due to an approximation whereby the matrix elements are evaluated by considering only the average of the diagonal baryon spin transitions. Consequently, our results should be regarded as averaged values for the masses and widths of the possible spin states that might eventually be resolved experimentally. At present, according to the Review of Particle Physics (RPP)~\cite{ParticleDataGroup:2024cfk}, the spin and parity of the $\Omega(2380)$ resonance have not yet been determined.

\section{Results}
\label{Sec:Res}
We first solve the Bethe--Salpeter equation with the potential given in Eq.~\eqref{eq:V}. Employing a cutoff regularization for the loop function \(G\) introduces a parameter \(q_{\mathrm{max}}\), which sets the momentum-space range of the interaction. As a first step, we adjust \(q_{\mathrm{max}}\) to produce a bound state around $2380$~MeV, corresponding to the mass of \(\Omega(2380)\). This serves as an initial consistency check of the theoretical framework.

The appearance of a bound state is nontrivial since the diagonal elements of the potential in Eq.~\eqref{eq:C} vanish. Similarly to the case of \(\Omega(2012)\), the bound state arises solely due to coupled-channel dynamics. Moreover, the value of \(q_{\mathrm{max}}\) should be physically reasonable, i.e., of the order of the vector-meson mass, as expected for the range of the interaction in momentum space~\cite{Gamermann:2009uq,Song:2022yvz}.

At this point, we find instructive to recall why the coupled channels can generate a bound state even when the diagonal potentials in those channels vanish (see Eqs.~\eqref{eq:V} and~\eqref{eq:C}). This is best illustrated in a two-channel system (1,2), with interaction matrix elements $V_{11}, V_{12}=V_{21}, V_{22}$ and diagonal loop functions $G_1, G_2$. Eliminating, for instance, channel 2, the scattering amplitude for the remaining single channel can be written in terms of an effective potential~\cite{Hyodo:2013nka,Wang:2022pin,Aceti:2014ala}
\begin{equation}
V_{\text{eff}}=V_{11}+\frac{V_{12}^2G_2}{(1-V_{22}G_2)}    \, .
\end{equation}
In a case such as the present one, where $V_{11} = V_{22} = 0$, the effective potential reduces to $V_{\text{eff}}=V_{12}^2G_2$. Since $V_{12}^2 > 0$ and $G_2 < 0$, the resulting effective interaction is attractive. Therefore, even in the absence of diagonal interactions, the coupled-channel dynamics can generate a bound state.

\begin{figure}[h!]  
    \centering
    \includegraphics[width=0.7\textwidth]{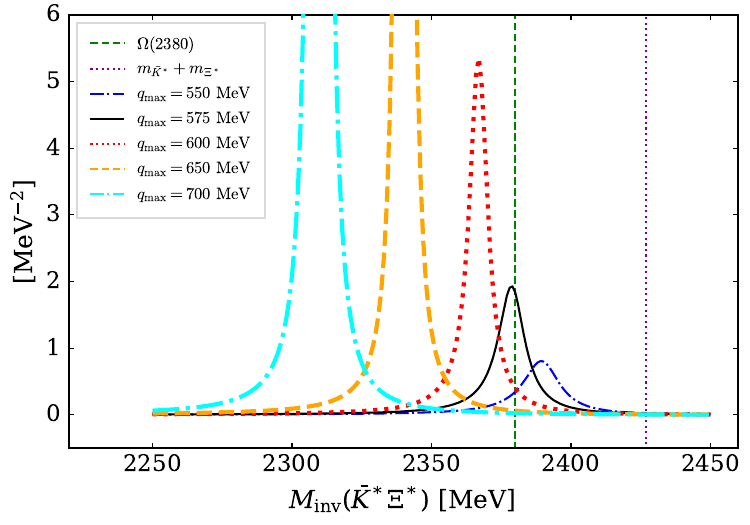}
    \caption{The modulus squared of the amplitude
$|T_{\bar{K}^*\Xi^*\to\bar{K}^*\Xi^*}|^2$
for different values of $q_{\mathrm{max}}$.
The left vertical (green dashed) line corresponds to 2380~MeV, while the right (purple dotted) line indicates the $\bar{K}^*\Xi^*$ threshold mass.}
    \label{fig:qmax}
\end{figure}

Fig.~\ref{fig:qmax} shows the modulus squared of the elastic amplitude $\bar{K}^*\Xi^*$, \( |T_{\bar{K}^*\Xi^*\to\bar{K}^*\Xi^*}|^2 \), for different values of \(q_{\mathrm{max}}\). Peaks corresponding to bound states are clearly observed in all cases. Based on these results, we select \(q_{\mathrm{max}} = 575\)~MeV, which produces a bound-state energy of $2380$~MeV. This value is then used in all subsequent calculations.

\begin{figure}[h!] 
    \centering
    \includegraphics[width=0.7\textwidth]{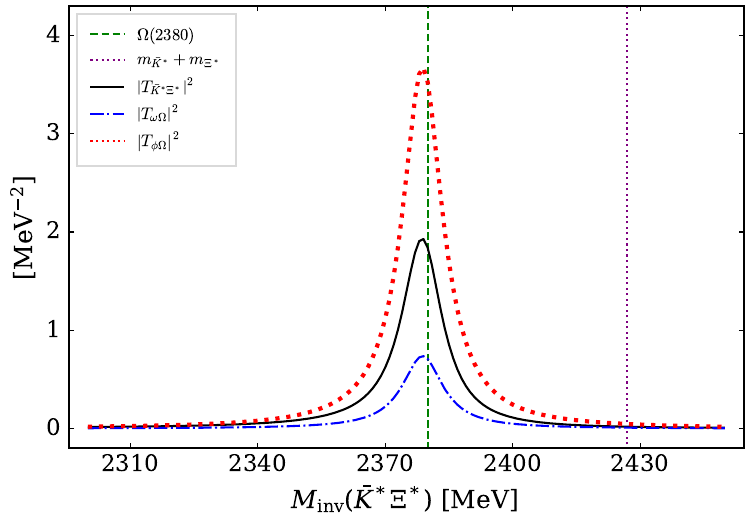}
    \caption{Modulus squared of the diagonal amplitudes $|T_{ii}|^2$ for the three channels. The vertical lines have the same meaning as in Fig.~\ref{fig:qmax}.}
    \label{fig:T}
\end{figure}

Fig.~\ref{fig:T} shows the modulus squared of the diagonal amplitudes, $|T_{ii}|^2$, for the three elastic channels: 
$\bar{K}^*\Xi^*$ (channel 1), $\omega\Omega$ (channel 2), and $\phi\Omega$ (channel 3).
The couplings of the pole to the different channels are defined as~\cite{Oset:2010tof}
\begin{equation}
|g_i|^2 = \frac{\Gamma}{2}
\sqrt{\left|T_{ii}(\sqrt{s}=M_R)\right|^2},
\qquad
g_j = g_i \frac{T_{ij}(\sqrt{s}=M_R)}{T_{ii}(\sqrt{s}=M_R)},
\end{equation}
where $M_R$ denotes the peak position of $|T_{ii}|^2$ and $\Gamma$ its full width at half maximum.
The wave function at the origin, which is relevant for short-range processes, is given by
\begin{equation}
\Psi_i \sim g_i\, G_i(\sqrt{s}=M_R).
\end{equation}

The resulting couplings and wave functions at the origin are shown in Table~\ref{tab:V1}.
The largest value of $|g_i G_i|$ corresponds to the $\bar{K}^*\Xi^*$ channel, indicating that this component dominates the structure of the $\Omega(2380)$ state.

\begin{table}[h!]
\centering
 \caption{Moduli of the coupling constants ($|g_i|$) and wave functions at the origin ($g_iG_i$ in MeV) of the three channels.}
 \label{tab:V1}
\setlength{\tabcolsep}{6.5pt}
\begin{tabular}{c|cc}
\hline
\hline
          ~           & ~$|g_i|$~ & ~$g_iG_i$~  \\
\hline
$\bar{K}^*\Xi^*$~       &        $2.84$         & $-27.79-i2.01$  \\
$ \omega\Omega$~   &        $2.25$         &            $-18.72$              \\
$ \phi\Omega$~    &        $3.36$         &     $11.31$                 \\

\hline
\hline
\end{tabular}
\end{table}

Next, we include the contributions from the box diagrams shown in Figs.~\ref{fig:KXistar} and~\ref{fig:KstarXi}. The coupled-channel space and the real part of the potential $V_{ij}$ remain unchanged,
while the imaginary parts of the box diagrams are added to the $\bar{K}^*\Xi^*$ elastic channel,
\begin{equation}
V_{11} \to V_{11} + i\,\mathrm{Im}\,\tilde{V}.
\end{equation}
We consider three different scenarios:
(i) only $\bar{K}\Xi^*$ box diagrams,
$V_{11} = i\,\mathrm{Im}\,\tilde{V}_{\bar{K}\Xi^*}$;
(ii) only $\bar{K}^*\Xi$ box diagrams,
$V_{11} = i\,\mathrm{Im}\,\tilde{V}_{\bar{K}^*\Xi}$;
and (iii) the inclusion of both contributions,
$V_{11} = i\,\mathrm{Im}\,\tilde{V}_{\bar{K}\Xi^*}
+ i\,\mathrm{Im}\,\tilde{V}_{\bar{K}^*\Xi}$.

\begin{figure}[h!] 
    \centering
    \includegraphics[width=0.75\textwidth]{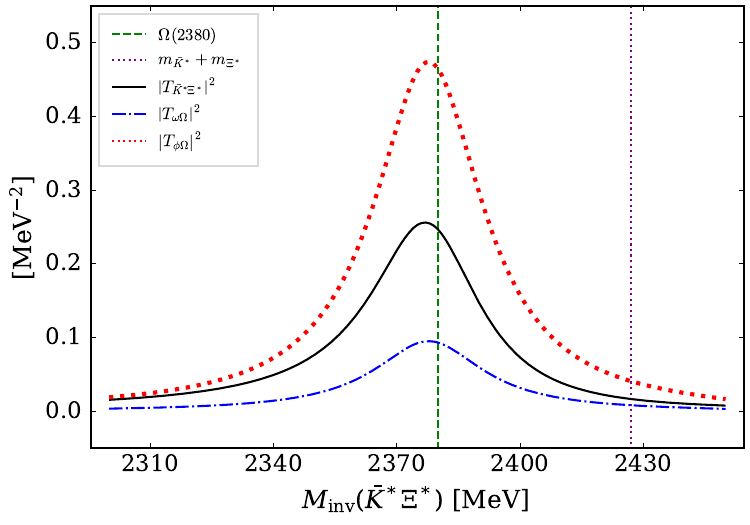}
    \caption{Same as Fig.~\ref{fig:T}, but considering the $\bar{K}\Xi^*$ box diagrams contributions.}
    \label{fig:ImK}
\end{figure}

\begin{figure}[h!] 
    \centering
    \includegraphics[width=0.7\textwidth]{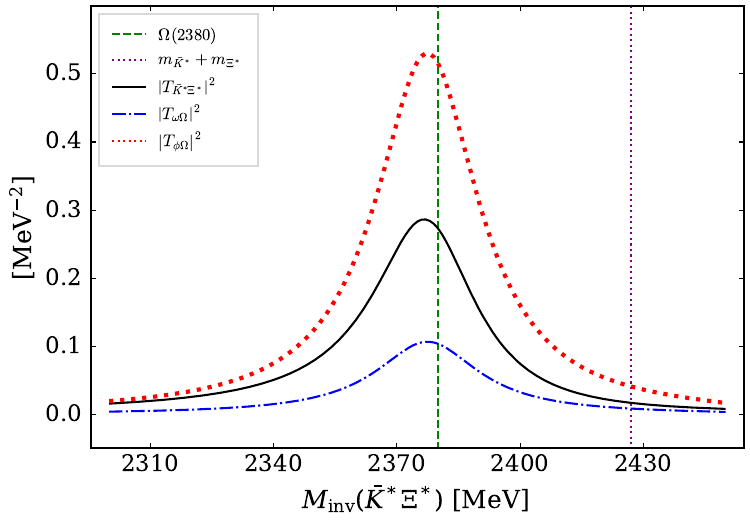}
    \caption{Same as Fig.~\ref{fig:T}, but considering the $\bar{K}^*\Xi$ box diagrams contributions.}
    \label{fig:ImKstar}
\end{figure}

\begin{figure}[h!] 
    \centering
    \includegraphics[width=0.7\textwidth]{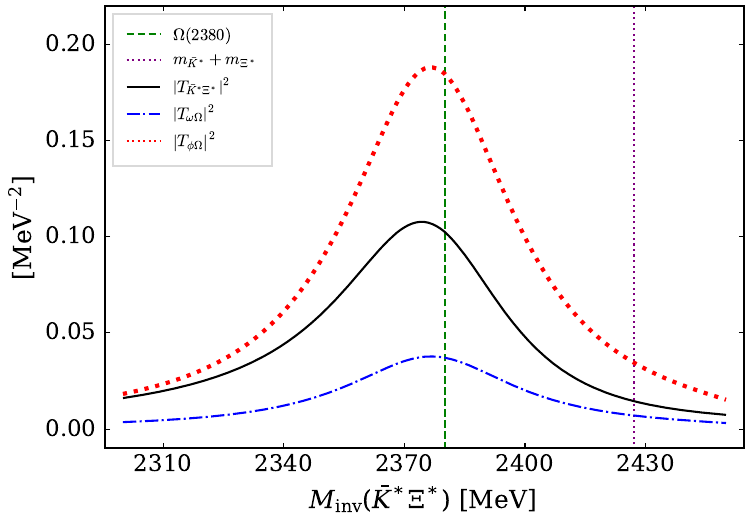}
    \caption{Same as Fig.~\ref{fig:T}, but considering the contributions of both kinds of box diagrams.}
    \label{fig:Imboth}
\end{figure}

From Figs.~\ref{fig:ImK}--\ref{fig:Imboth}, one observes that the peak positions remain close to
2380~MeV in all cases, with small downward shifts of about $2$--$4$~MeV compared to the result
without box diagrams shown in Fig.~\ref{fig:T}. In contrast, the widths increase substantially due to the opening of additional decay channels associated with the box-diagram contributions.

The corresponding couplings and wave functions at the origin ($g_i G_i$) for the three scenarios are collected in Table~\ref{tab:V2}. In all cases, the $\bar{K}^*\Xi^*$ component remains dominant, indicating that the internal structure of the $\Omega(2380)$ is only weakly affected by the inclusion of the box diagrams. It can be also appreciated that the values of the couplings and the wave functions at the origin do not change appreciably when the box diagrams are taken into account.

\begin{table}[h!]
\centering
 \caption{Same results with Table~\ref{tab:V1}, but only considering the $\bar{K}\Xi^*$ box diagrams in the left column, only considering the $\bar{K}^*\Xi$ box diagrams in the middle column, and considering both box diagrams in the right column.}
 \label{tab:V2}
\setlength{\tabcolsep}{6.5pt}
\begin{tabular}{c|cc|cc|cc}
\hline
\hline
~& \multicolumn{2}{c|}{$\bar{K}\Xi^*$ box diagrams} & \multicolumn{2}{c|}{$\bar{K}^*\Xi$ box diagrams} & \multicolumn{2}{c}{both box diagrams}\\
\cline{2-3}\cline{4-5}\cline{6-7}
          ~           & ~$|g_i|$~ & ~$g_iG_i$~ & ~$|g_i|$~ & ~$g_iG_i$~ & ~$|g_i|$~ & ~$g_iG_i$~ \\
\hline
$\bar{K}^*\Xi^*$~       &        $2.88$         & $-27.77-i1.88$   &        $2.84$         & $-24.45-i1.88$ &        $2.91$         & $-27.73-i1.77$  \\
$ \omega\Omega$~   &        $2.28$         &            $-18.61+i2.48$     &        $2.25$         &            $-18.42+i2.24$   &        $2.26$         &            $-17.92+i4.61$                \\
$ \phi\Omega$~    &        $3.40$         &     $11.32-i1.51$          &        $3.36$         &     $11.20-i1.36$         &        $3.38$         &     $10.96-i2.82$                   \\

\hline
\hline
\end{tabular}
\end{table}

\begin{table}[h!]
\centering
 \caption{The peak positions and widths of four different cases, without considering any box diagram (the original case), only consider the $\bar{K}\Xi^*$ box diagrams, only consider the $\bar{K}^*\Xi$ box diagrams and consider both kinds of box diagrams. The left column corresponds to the results when we take $\Lambda=1$~GeV, while for the right one we take $\Lambda=800$~MeV. In all cases, the widths of the $\bar{K}^*$ and $\Xi^*$ are considered.}
 \label{tab:width}
\setlength{\tabcolsep}{6.5pt}
\begin{tabular}{c|cc|cc}
\hline
\hline
         & \multicolumn{2}{c|}{$\Lambda=1$~GeV} & \multicolumn{2}{c}{$\Lambda=800$~MeV} \\
\cline{2-3}\cline{4-5}
          ~           & peak position  & width  & peak position  & width \\
\hline
Without box diagrams~       &        $2379.0$~MeV         & $\Gamma_1=11.6$~MeV  &        $2379.0$~MeV        & $\Gamma_1=11.6$~MeV\\
Only $\bar{K}\Xi^*$ box diagrams~       &        $2376.9$~MeV         & $\Gamma_2=32.5$~MeV  &        $2377.7$~MeV         & $\Gamma_2=26.5$~MeV  \\
Only $\bar{K}^*\Xi$ box diagrams~   &        $2377.2$~MeV         &            $\Gamma_3=30.0$~MeV    &        $2377.4$~MeV         &            $\Gamma_3=27.4$~MeV          \\
Both box diagrams~    &        $2375.2$~MeV         &     $\Gamma_4=51.5$~MeV   &        $2375.9$~MeV         &     $\Gamma_4=42.2$~MeV                      \\
\hline
\hline
\end{tabular}
\end{table}

Finally, we summarize in the left column of Table~\ref{tab:width} the peak positions and widths for the original case, as well as for the three scenarios including box-diagram contributions. 
In the original case, the width $\Gamma_1$ originates solely from the finite widths of the $\bar{K}^*$ and $\Xi^*$ particles, which are incorporated through the convolution of the loop functions $G$ in Eq.~\eqref{eq:G1}. 
For the other three cases, the total widths receive contributions both from this convolution and from the box diagrams.

The $\bar{K}\Xi^*$ box diagrams correspond to the $\bar{K}\Xi^*$ decay mode of the $\Omega(2380)$, while the $\bar{K}^*\Xi$ box diagrams represent the $\bar{K}^*\Xi$ decay mode. Including both types of box diagrams effectively accounts for the $\bar{K}\Xi\pi$ decay channel.
Therefore, the partial widths associated with these decay modes can be estimated by subtracting $\Gamma_1$ from the corresponding total widths,
\begin{align}
    \Gamma_{\bar{K}\Xi^*} &= \Gamma_2 - \Gamma_1 = 20.9~\textrm{MeV}, \label{eq:gamma2}
\\
    \Gamma_{\bar{K}^*\Xi} &= \Gamma_3 - \Gamma_1 = 18.4~\textrm{MeV}, \label{eq:gamma3}
\\
    \Gamma_{\bar{K}\Xi\pi} &= \Gamma_4 - \Gamma_1 = 39.9~\textrm{MeV}. \label{eq:gamma4}
\end{align}

From these results we also find the approximate relation $\Gamma_{\bar{K}\Xi^*}+\Gamma_{\bar{K}^*\Xi}\approx\Gamma_{\bar{K}\Xi\pi}$, which provides a nontrivial consistency check of our approach. The corresponding branching ratios are then given by
\begin{align}
    \textrm{Br}[{\bar{K}\Xi^*}] &=\Gamma_{\bar{K}\Xi^*}/\Gamma_{\bar{K}\Xi\pi}=0.53, \label{eq:br1}
\\
     \textrm{Br}[{\bar{K}^*\Xi}] &=\Gamma_{\bar{K}^*\Xi}/\Gamma_{\bar{K}\Xi\pi}=0.47. \label{eq:br2}
\end{align}
Comparing these results with those provided by PDG~\cite{ParticleDataGroup:2024cfk}, we can appreciate that our results are compatible with the ratio $\Gamma_{\bar{K}^*\Xi}/\Gamma_{\bar{K}\Xi\pi}$ within the errors. For the case of $\Gamma_{\bar{K}\Xi^*}/\Gamma_{\bar{K}\Xi\pi}$ we obtain a value of $0.53$, which lies close to the upper experimental bound. It would be interesting to have absolute values for this rate in the future, to allow for a better comparison. As for the total width, we obtain $\Gamma\approx51.5$ MeV. This should be compared with the experimental width of $\Gamma_{\textrm{exp}}=26\pm23$~MeV. Within uncertainties, these values are compatible. This result provides further support for the molecular interpretation of the $\Omega(2380)$, since the width depends only weakly on the cutoff $q_{\mathrm{max}}$, which is fixed by reproducing the physical mass of the state, and on the parameter $\Lambda$ associated with the off-shell behavior of the exchanged pion.

To estimate the associated theoretical uncertainties, we repeat the calculation using $\Lambda = 800$~MeV, a value commonly employed in the literature.
The corresponding results are shown in the right column of Table~\ref{tab:width}.
In this case, the total width decreases from $51.5$~MeV to $42.2$~MeV, well within the experimental range.
The partial decay widths change moderately, and the resulting branching ratios become
\begin{align}
    \mathrm{Br}[\bar{K}\Xi^*] &= 0.48, \label{eq:br3}
\\
    \mathrm{Br}[\bar{K}^*\Xi] &= 0.52. \label{eq:br4}
\end{align}

As seen from Eqs.~\eqref{eq:br3} and~\eqref{eq:br4}, the ratio
$\Gamma_{\bar{K}\Xi^*}/\Gamma_{\bar{K}\Xi\pi}$ becomes smaller than
$\Gamma_{\bar{K}^*\Xi}/\Gamma_{\bar{K}\Xi\pi}$, in contrast to the previous case, and closer to the experimental bound.
Overall, taking into account the theoretical uncertainties, the predicted total width and partial decay widths show a remarkable agreement with the available experimental information.

\section{Conclusions} 
\label{Sec:Conclu}

The discovery of the $\Omega(2012)$ triggered extensive discussion regarding its nature. Measurements of its decay into $\bar{K}\Xi$ and $\bar{K}\pi\Xi$, together with different theoretical studies, led the Belle Collaboration to favor a molecular interpretation generated from the $\bar{K}\Xi^*$ and $\eta\Omega$ interactions, with $\bar{K}\Xi$ as a decay channel. It is worth noting that such a state had been predicted prior to its experimental observation \cite{Sarkar:2004jh}. These ideas were subsequently extended to the $\bar{K}^*\Xi^*$, $\omega\Omega$, and $\phi\Omega$ interactions, where a spin-degenerate negative-parity resonance was predicted within an $s$-wave, spin-independent framework \cite{Sarkar:2010saz}.

In the present work, we identify the $\Omega(2380)$ as a candidate for this predicted spin-degenerate resonance, with spin–parity assignments $J^P=1/2^-$, $3/2^-$ and $5/2^-$, and extend the approach of Ref.~\cite{Sarkar:2010saz} by incorporating the $\bar{K}\Xi^*$ and $\bar{K}^*\Xi$ decay channels through box-diagram mechanisms. These contributions provide a sizable enhancement to the width compared to the natural decay of the $\bar{K}^*\Xi^*$ component. The regulator of the loop functions was fixed to reproduce the mass of the $\Omega(2380)$, yielding a cutoff value well within the natural range expected from light vector-meson exchange. The resulting total width, as well as the partial decay widths into the $\bar{K}\Xi^*$ and $\bar{K}^*\Xi$ channels, are found to be compatible with the available experimental information.

Overall, our results support the interpretation of the $\Omega(2380)$ as a dynamically generated resonance arising from vector meson-decuplet baryon interactions. Further experimental and theoretical studies are desirable to validate this picture. In particular, femtoscopic measurements of correlation functions involving the relevant channels—most notably the $\bar{K}^*\Xi^*$ system—could provide valuable constraints, following dedicated theoretical analyses along the lines of Refs.~\cite{Sarti:2023wlg,Feijoo:2024bvn,Encarnacion:2024jge,Albaladejo:2025lhn} applied to recent ALICE data \cite{ALICE:2022yyh,ALICE:2023wjz,ALICE:2021cpv,ALICE:2025flv,ALICE:2023eyl}. Alternatively, complementary insight could be obtained through model-independent inverse-amplitude analyses, as proposed in Refs.~\cite{Albaladejo:2023wmv,Feijoo:2023sfe}.

\vspace{15pt}

\begin{acknowledgements}
This work is supported by the Spanish Ministerio de Ciencia e Innovaci\'on (MICINN) under contracts PID2020-112777GB-I00, PID2023-147458NB-C21 and CEX2023-001292-S; by Generalitat Valenciana under contracts PROMETEO/2020/023 and  CIPROM/2023/59. Y.-Y. L. is supported in part by the Guangdong Provincial international exchange program for outstanding young talents of scientific research in 2024. E. O. and A. F. thank the warm support of the ACVJLI.

\end{acknowledgements}

\clearpage
\bibliography{2380Ref}

@article{ParticleDataGroup:2018ovx,
    author = "Tanabashi, M. and others",
    collaboration = "Particle Data Group",
    title = "{Review of Particle Physics}",
    doi = "10.1103/PhysRevD.98.030001",
    journal = "Phys. Rev. D",
    volume = "98",
    number = "3",
    pages = "030001",
    year = "2018"
}

@article{Capstick:1986ter,
    author = "Capstick, Simon and Isgur, Nathan",
    title = "{Baryons in a relativized quark model with chromodynamics}",
    doi = "10.1103/physrevd.34.2809",
    journal = "Phys. Rev. D",
    volume = "34",
    number = "9",
    pages = "2809--2835",
    year = "1986"
}

@article{Loring:2001ky,
    author = "Loring, Ulrich and Metsch, Bernard C. and Petry, Herbert R.",
    title = "{The Light baryon spectrum in a relativistic quark model with instanton induced quark forces: The Strange baryon spectrum}",
    eprint = "hep-ph/0103290",
    archivePrefix = "arXiv",
    reportNumber = "TK-01-07",
    doi = "10.1007/s100500170106",
    journal = "Eur. Phys. J. A",
    volume = "10",
    pages = "447--486",
    year = "2001"
}

@article{Pervin:2007wa,
    author = "Pervin, Muslema and Roberts, Winston",
    title = "{Strangeness -2 and -3 baryons in a constituent quark model}",
    eprint = "0709.4000",
    archivePrefix = "arXiv",
    primaryClass = "nucl-th",
    reportNumber = "JLAB-THY-07-728",
    doi = "10.1103/PhysRevC.77.025202",
    journal = "Phys. Rev. C",
    volume = "77",
    pages = "025202",
    year = "2008"
}

@article{Faustov:2015eba,
    author = "Faustov, R. N. and Galkin, V. O.",
    title = "{Strange baryon spectroscopy in the relativistic quark model}",
    eprint = "1507.04530",
    archivePrefix = "arXiv",
    primaryClass = "hep-ph",
    doi = "10.1103/PhysRevD.92.054005",
    journal = "Phys. Rev. D",
    volume = "92",
    number = "5",
    pages = "054005",
    year = "2015"
}

@article{Chao:1980em,
    author = "Chao, Kuang-Ta and Isgur, Nathan and Karl, Gabriel",
    title = "{Strangeness -2 and -3 Baryons in a Quark Model With Chromodynamics}",
    reportNumber = "Print-80-0704 (TORONTO)",
    doi = "10.1103/PhysRevD.23.155",
    journal = "Phys. Rev. D",
    volume = "23",
    pages = "155",
    year = "1981"
}

@article{Kalman:1982ut,
    author = "Kalman, Calvin S.",
    title = "{$P$ Wave Baryons in a Consistent Quark Model With Hyperfine Interactions}",
    reportNumber = "CUQ/EPP-40",
    doi = "10.1103/PhysRevD.26.2326",
    journal = "Phys. Rev. D",
    volume = "26",
    pages = "2326",
    year = "1982"
}

@article{An:2013zoa,
    author = "An, C. S. and Metsch, B. Ch. and Zou, B. S.",
    title = "{Mixing of the low-lying three- and five-quark $\Omega$ states with negative parity}",
    eprint = "1304.6046",
    archivePrefix = "arXiv",
    primaryClass = "hep-ph",
    doi = "10.1103/PhysRevC.87.065207",
    journal = "Phys. Rev. C",
    volume = "87",
    number = "6",
    pages = "065207",
    year = "2013"
}

@article{An:2014lga,
    author = "An, C. S. and Zou, B. S.",
    title = "{Low-lying $\Omega$ states with negative parity in an extended quark model with Nambu-Jona-Lasinio interaction}",
    eprint = "1403.7897",
    archivePrefix = "arXiv",
    primaryClass = "hep-ph",
    doi = "10.1103/PhysRevC.89.055209",
    journal = "Phys. Rev. C",
    volume = "89",
    number = "5",
    pages = "055209",
    year = "2014"
}

@article{Oh:2007cr,
    author = "Oh, Yongseok",
    title = "{Xi and Omega baryons in the Skyrme model}",
    eprint = "hep-ph/0702126",
    archivePrefix = "arXiv",
    doi = "10.1103/PhysRevD.75.074002",
    journal = "Phys. Rev. D",
    volume = "75",
    pages = "074002",
    year = "2007"
}

@article{Engel:2013ig,
    author = {Engel, Georg P. and Lang, C. B. and Mohler, Daniel and Sch{\"a}fer, Andreas},
    collaboration = "BGR",
    title = "{QCD with Two Light Dynamical Chirally Improved Quarks: Baryons}",
    eprint = "1301.4318",
    archivePrefix = "arXiv",
    primaryClass = "hep-lat",
    reportNumber = "FERMILAB-PUB-13-028-T",
    doi = "10.1103/PhysRevD.87.074504",
    journal = "Phys. Rev. D",
    volume = "87",
    number = "7",
    pages = "074504",
    year = "2013"
}

@article{CLQCD:2015bgi,
    author = "Liang, Jian and Sun, Wei and Chen, Ying and Qiu, Wei-Feng and Gong, Ming and Liu, Chuan and Liu, Yu-Bin and Liu, Zhaofeng and Ma, Jian-Ping and Zhang, Jian-Bo",
    collaboration = "CLQCD",
    title = "{Spectrum and Bethe-Salpeter amplitudes of $\Omega$ baryons from lattice QCD}",
    eprint = "1511.04294",
    archivePrefix = "arXiv",
    primaryClass = "hep-lat",
    doi = "10.1088/1674-1137/40/4/041001",
    journal = "Chin. Phys. C",
    volume = "40",
    number = "4",
    pages = "041001",
    year = "2016"
}

@article{Belle:2018mqs,
    author = "Yelton, J. and others",
    collaboration = "Belle",
    title = "{Observation of an Excited $\Omega^-$ Baryon}",
    eprint = "1805.09384",
    archivePrefix = "arXiv",
    primaryClass = "hep-ex",
    reportNumber = "BELLE-PREPRINT-2018-09, KEK-PREPRINT-2018-3, Belle Preprint 2018-09, KEK Preprint 2018-3",
    doi = "10.1103/PhysRevLett.121.052003",
    journal = "Phys. Rev. Lett.",
    volume = "121",
    number = "5",
    pages = "052003",
    year = "2018"
}

@article{Belle:2019zco,
    author = "Jia, S. and others",
    collaboration = "Belle",
    title = "{Search for $\Omega(2012)\to K\Xi(1530) \to K\pi\Xi$ at Belle}",
    eprint = "1906.00194",
    archivePrefix = "arXiv",
    primaryClass = "hep-ex",
    reportNumber = "Belle Preprint {\#} 2019-10, and KEK Preprint {\#}: 2019-8, Belle Preprint-2019-10, KEK Preprint-2019-8",
    doi = "10.1103/PhysRevD.100.032006",
    journal = "Phys. Rev. D",
    volume = "100",
    number = "3",
    pages = "032006",
    year = "2019"
}

@article{Belle:2022mrg,
    author = "Jia, S. and others",
    collaboration = "Belle",
    title = "{Observation of {\ensuremath{\Omega}}(2012){\ensuremath{-}}{\textrightarrow}{\ensuremath{\Xi}}(1530)K{\textasciimacron} and measurement of the effective couplings of {\ensuremath{\Omega}}(2012){\ensuremath{-}} to {\ensuremath{\Xi}}(1530)K{\textasciimacron} and {\ensuremath{\Xi}}K{\textasciimacron}}",
    eprint = "2207.03090",
    archivePrefix = "arXiv",
    primaryClass = "hep-ex",
    reportNumber = "Belle Preprint 2024-08; KEK Preprint 2024-25, Belle Preprint 2022-14; KEK Preprint 2022-14",
    doi = "10.1016/j.physletb.2024.139224",
    journal = "Phys. Lett. B",
    volume = "860",
    pages = "139224",
    year = "2025"
}

@article{BESIII:2024eqk,
    author = "Ablikim, Medina and others",
    collaboration = "BESIII",
    title = "{Evidence for Two Excited {\ensuremath{\Omega}}- Hyperons}",
    eprint = "2411.11648",
    archivePrefix = "arXiv",
    primaryClass = "hep-ex",
    doi = "10.1103/PhysRevLett.134.131903",
    journal = "Phys. Rev. Lett.",
    volume = "134",
    number = "13",
    pages = "131903",
    year = "2025"
}

@article{ALICE:2025atb,
    author = "Acharya, Shreyasi and others",
    collaboration = "ALICE",
    title = "{Observation of the {\ensuremath{\Omega}}(2012) baryon at the LHC}",
    eprint = "2502.18063",
    archivePrefix = "arXiv",
    primaryClass = "hep-ex",
    reportNumber = "CERN-EP-2025-027",
    doi = "10.1103/v4mh-3r8z",
    journal = "Phys. Rev. D",
    volume = "112",
    number = "9",
    pages = "092002",
    year = "2025"
}

@article{Xiao:2018pwe,
    author = "Xiao, Li-Ye and Zhong, Xian-Hui",
    title = "{Possible interpretation of the newly observed $\Omega$(2012) state}",
    eprint = "1805.11285",
    archivePrefix = "arXiv",
    primaryClass = "hep-ph",
    doi = "10.1103/PhysRevD.98.034004",
    journal = "Phys. Rev. D",
    volume = "98",
    number = "3",
    pages = "034004",
    year = "2018"
}

@article{Aliev:2018syi,
    author = "Aliev, T. M. and Azizi, K. and Sarac, Y. and Sundu, H.",
    title = "{Interpretation of the newly discovered $\Omega$(2012)}",
    eprint = "1806.01626",
    archivePrefix = "arXiv",
    primaryClass = "hep-ph",
    doi = "10.1103/PhysRevD.98.014031",
    journal = "Phys. Rev. D",
    volume = "98",
    number = "1",
    pages = "014031",
    year = "2018"
}

@article{Aliev:2018yjo,
    author = "Aliev, T. M. and Azizi, K. and Sarac, Y. and Sundu, H.",
    title = "{Nature of the $\Omega (2012)$ through its strong decays}",
    eprint = "1807.02145",
    archivePrefix = "arXiv",
    primaryClass = "hep-ph",
    doi = "10.1140/epjc/s10052-018-6375-y",
    journal = "Eur. Phys. J. C",
    volume = "78",
    number = "11",
    pages = "894",
    year = "2018"
}

@article{Wang:2018hmi,
    author = {Wang, Zuo-Yun and Gui, Long-Cheng and L{\"u}, Qi-Fang and Xiao, Li-Ye and Zhong, Xian-Hui},
    title = "{Newly observed $\Omega(2012)$ state and strong decays of the low-lying $\Omega$ excitations}",
    eprint = "1810.08318",
    archivePrefix = "arXiv",
    primaryClass = "hep-ph",
    doi = "10.1103/PhysRevD.98.114023",
    journal = "Phys. Rev. D",
    volume = "98",
    number = "11",
    pages = "114023",
    year = "2018"
}

@article{Polyakov:2018mow,
    author = "Polyakov, Maxim V. and Son, Hyeon-Dong and Sun, Bao-Dong and Tandogan, Asli",
    title = "{{\ensuremath{\Omega}}(2012) through the looking glass of flavour SU (3)}",
    eprint = "1806.04427",
    archivePrefix = "arXiv",
    primaryClass = "hep-ph",
    doi = "10.1016/j.physletb.2019.03.054",
    journal = "Phys. Lett. B",
    volume = "792",
    pages = "315--319",
    year = "2019"
}

@article{Liu:2019wdr,
    author = {Liu, Ming-Sheng and Wang, Kai-Lei and L{\"u}, Qi-Fang and Zhong, Xian-Hui},
    title = "{$\Omega$ baryon spectrum and their decays in a constituent quark model}",
    eprint = "1910.10322",
    archivePrefix = "arXiv",
    primaryClass = "hep-ph",
    doi = "10.1103/PhysRevD.101.016002",
    journal = "Phys. Rev. D",
    volume = "101",
    number = "1",
    pages = "016002",
    year = "2020"
}

@article{Liu:2020yen,
    author = "Liu, Xuejie and Huang, Hongxia and Ping, Jialun and Chen, Dianyong",
    title = "{Investigating $\Omega(2012)$ as a molecular state}",
    eprint = "2010.15398",
    archivePrefix = "arXiv",
    primaryClass = "hep-ph",
    doi = "10.1103/PhysRevC.103.025202",
    journal = "Phys. Rev. C",
    volume = "103",
    number = "2",
    pages = "025202",
    year = "2021"
}

@article{Arifi:2022ntc,
    author = "Arifi, Ahmad Jafar and Suenaga, Daiki and Hosaka, Atsushi and Oh, Yongseok",
    title = "{Strong decays of multistrangeness baryon resonances in the quark model}",
    eprint = "2201.10427",
    archivePrefix = "arXiv",
    primaryClass = "hep-ph",
    doi = "10.1103/PhysRevD.105.094006",
    journal = "Phys. Rev. D",
    volume = "105",
    number = "9",
    pages = "094006",
    year = "2022"
}

@article{Zhong:2022cjx,
    author = "Zhong, Hui-Hua and Ni, Ru-Hui and Chen, Mu-Yang and Zhong, Xian-Hui and Xie, Ju-Jun",
    title = "{Further study of within a chiral quark model*}",
    eprint = "2209.09398",
    archivePrefix = "arXiv",
    primaryClass = "hep-ph",
    doi = "10.1088/1674-1137/acc9a2",
    journal = "Chin. Phys. C",
    volume = "47",
    number = "6",
    pages = "063104",
    year = "2023"
}

@article{Wang:2022zja,
    author = {Wang, Kai-Lei and L{\"u}, Qi-Fang and Xie, Ju-Jun and Zhong, Xian-Hui},
    title = "{Toward discovering the excited {\ensuremath{\Omega}} baryons through nonleptonic weak decays of {\ensuremath{\Omega}}c}",
    eprint = "2203.04458",
    archivePrefix = "arXiv",
    primaryClass = "hep-ph",
    doi = "10.1103/PhysRevD.107.034015",
    journal = "Phys. Rev. D",
    volume = "107",
    number = "3",
    pages = "034015",
    year = "2023"
}

@article{Luo:2025cqs,
    author = "Luo, Si-Qiang and Liu, Xiang",
    title = "{Identifying triple-strangeness {\ensuremath{\Omega}} hyperons in light of recent experimental results}",
    eprint = "2504.14648",
    archivePrefix = "arXiv",
    primaryClass = "hep-ph",
    doi = "10.1103/18md-j4bf",
    journal = "Phys. Rev. D",
    volume = "112",
    number = "1",
    pages = "014047",
    year = "2025"
}

@article{Valderrama:2018bmv,
    author = "Valderrama, M. Pavon",
    title = "{$\Omega(2012)$ as a hadronic molecule}",
    eprint = "1807.00718",
    archivePrefix = "arXiv",
    primaryClass = "hep-ph",
    doi = "10.1103/PhysRevD.98.054009",
    journal = "Phys. Rev. D",
    volume = "98",
    number = "5",
    pages = "054009",
    year = "2018"
}

@article{Lin:2018nqd,
    author = "Lin, Yong-Hui and Zou, Bing-Song",
    title = "{Hadronic molecular assignment for the newly observed $\Omega^*$ state}",
    eprint = "1807.00997",
    archivePrefix = "arXiv",
    primaryClass = "hep-ph",
    doi = "10.1103/PhysRevD.98.056013",
    journal = "Phys. Rev. D",
    volume = "98",
    number = "5",
    pages = "056013",
    year = "2018"
}

@article{Huang:2018wth,
    author = "Huang, Yin and Liu, Ming-Zhu and Lu, Jun-Xu and Xie, Ju-Jun and Geng, Li-Sheng",
    title = "{Strong decay modes $\bar{K}\Xi$ and $\bar{K}\Xi\pi$ of the $\Omega(2012)$ in the $\bar{K}\Xi(1530)$ and $\eta\Omega$ molecular scenario}",
    eprint = "1807.06485",
    archivePrefix = "arXiv",
    primaryClass = "hep-ph",
    doi = "10.1103/PhysRevD.98.076012",
    journal = "Phys. Rev. D",
    volume = "98",
    number = "7",
    pages = "076012",
    year = "2018"
}

@article{Gutsche:2019eoh,
    author = "Gutsche, Thomas and Lyubovitskij, Valery E.",
    title = "{Strong decays of the hadronic molecule $\Omega^\ast (2012)$}",
    eprint = "1912.10894",
    archivePrefix = "arXiv",
    primaryClass = "hep-ph",
    doi = "10.1088/1361-6471/abcb9f",
    journal = "J. Phys. G",
    volume = "48",
    number = "2",
    pages = "025001",
    year = "2020"
}

@article{Lin:2019tex,
    author = "Lin, Yonghui and Lin, Yong-Hui and Wang, Fei and Zou, Bingsong and Zou, Bing-Song",
    title = "{Reanalysis of the newly observed $\Omega^*$ state in a hadronic molecule model}",
    eprint = "1910.13919",
    archivePrefix = "arXiv",
    primaryClass = "hep-ph",
    doi = "10.1103/PhysRevD.102.074025",
    journal = "Phys. Rev. D",
    volume = "102",
    number = "7",
    pages = "074025",
    year = "2020"
}

@article{Lu:2022puv,
    author = {L{\"u}, Qi-Fang and Nagahiro, Hideko and Hosaka, Atsushi},
    title = "{Understanding the nature of {\ensuremath{\Omega}}(2012) in a coupled-channel approach}",
    eprint = "2212.02783",
    archivePrefix = "arXiv",
    primaryClass = "hep-ph",
    doi = "10.1103/PhysRevD.107.014025",
    journal = "Phys. Rev. D",
    volume = "107",
    number = "1",
    pages = "014025",
    year = "2023"
}

@article{Lu:2020ste,
    author = "Lu, Jun-Xu and Zeng, Chun-Hua and Wang, En and Xie, Ju-Jun and Geng, Li-Sheng",
    title = "{Revisiting the $\Omega(2012)$ as a hadronic molecule and its strong decays}",
    eprint = "2003.07588",
    archivePrefix = "arXiv",
    primaryClass = "hep-ph",
    doi = "10.1140/epjc/s10052-020-7944-4",
    journal = "Eur. Phys. J. C",
    volume = "80",
    number = "5",
    pages = "361",
    year = "2020"
}

@article{Ikeno:2020vqv,
    author = "Ikeno, Natsumi and Toledo, Genaro and Oset, Eulogio",
    title = "{Molecular picture for the $\Omega(2012)$  revisited}",
    eprint = "2003.07580",
    archivePrefix = "arXiv",
    primaryClass = "hep-ph",
    doi = "10.1103/PhysRevD.101.094016",
    journal = "Phys. Rev. D",
    volume = "101",
    number = "9",
    pages = "094016",
    year = "2020"
}

@article{Klingl:1997kf,
    author = "Klingl, F. and Kaiser, Norbert and Weise, W.",
    title = "{Current correlation functions, QCD sum rules and vector mesons in baryonic matter}",
    eprint = "hep-ph/9704398",
    archivePrefix = "arXiv",
    doi = "10.1016/S0375-9474(97)88960-9",
    journal = "Nucl. Phys. A",
    volume = "624",
    pages = "527--563",
    year = "1997"
}

@article{Palomar:2002hk,
    author = "Palomar, J. E. and Oset, E.",
    title = "{The phi NN coupling from chiral loops}",
    eprint = "nucl-th/0208013",
    archivePrefix = "arXiv",
    reportNumber = "FTUV-02-0808, IFIC-02-0808",
    doi = "10.1016/S0375-9474(02)01405-7",
    journal = "Nucl. Phys. A",
    volume = "716",
    pages = "169--185",
    year = "2003"
}

@article{Oset:2010tof,
    author = "Oset, E. and Ramos, A.",
    title = "{Dynamically generated resonances from the vector octet-baryon octet interaction}",
    eprint = "0905.0973",
    archivePrefix = "arXiv",
    primaryClass = "hep-ph",
    doi = "10.1140/epja/i2010-10957-3",
    journal = "Eur. Phys. J. A",
    volume = "44",
    pages = "445--454",
    year = "2010"
}

@article{Ikeno:2022jpe,
    author = "Ikeno, Natsumi and Liang, Wei-Hong and Toledo, Genaro and Oset, Eulogio",
    title = "{Interpretation of the \ensuremath{\Omega}c\textrightarrow{}\ensuremath{\pi^+}\ensuremath{\Omega}(2012)\textrightarrow{}\ensuremath{\pi^+}(K\textasciimacron{}\ensuremath{\Xi}) relative to \ensuremath{\Omega}c\textrightarrow{}\ensuremath{\pi^+}K\textasciimacron{}\ensuremath{\Xi} from the \ensuremath{\Omega}(2012) molecular perspective}",
    eprint = "2204.13396",
    archivePrefix = "arXiv",
    primaryClass = "hep-ph",
    doi = "10.1103/PhysRevD.106.034022",
    journal = "Phys. Rev. D",
    volume = "106",
    number = "3",
    pages = "034022",
    year = "2022"
}

@article{Han:2025gkp,
    author = "Han, Fang-Chao and Liu, Zhan-Wei and Leinweber, Derek B. and Thomas, Anthony W.",
    title = "{Structure of the {\ensuremath{\Omega}}-(2012) with Hamiltonian effective field theory}",
    eprint = "2507.06682",
    archivePrefix = "arXiv",
    primaryClass = "hep-ph",
    reportNumber = "ADP-25-25/T1287",
    doi = "10.1103/xdbd-v5hb",
    journal = "Phys. Rev. D",
    volume = "112",
    number = "5",
    pages = "L051503",
    year = "2025"
}

@article{Shen:2025xcq,
    author = "Shen, Qing-Hua and Lu, Jun-Xu and Geng, Li-Sheng and Liu, Xiang and Xie, Ju-Jun",
    title = "{Radiative decays of the $Ω(2012)$ as a hadronic molecule}",
    eprint = "2510.13623",
    archivePrefix = "arXiv",
    journal = {arXiv preprint},
    primaryClass = "hep-ph",
    month = "10",
    year = "2025"
}

@article{Jenkins:1991es,
    author = "Jenkins, Elizabeth Ellen and Manohar, Aneesh V.",
    title = "{Chiral corrections to the baryon axial currents}",
    reportNumber = "UCSD-PTH-91-05",
    doi = "10.1016/0370-2693(91)90840-M",
    journal = "Phys. Lett. B",
    volume = "259",
    pages = "353--358",
    year = "1991"
}

@article{Kolomeitsev:2003kt,
    author = "Kolomeitsev, E. E. and Lutz, M. F. M.",
    title = "{On baryon resonances and chiral symmetry}",
    eprint = "nucl-th/0305101",
    archivePrefix = "arXiv",
    reportNumber = "GSI-PREPRINT-2003-17",
    doi = "10.1016/j.physletb.2004.01.066",
    journal = "Phys. Lett. B",
    volume = "585",
    pages = "243--252",
    year = "2004"
}

@article{Sarkar:2004jh,
    author = "Sarkar, Sourav and Oset, E. and Vicente Vacas, M. J.",
    title = "{Baryonic resonances from baryon decuplet-meson octet interaction}",
    eprint = "nucl-th/0407025",
    archivePrefix = "arXiv",
    reportNumber = "FTUV-04-0707, IFIC-04-0707",
    doi = "10.1016/j.nuclphysa.2005.01.006",
    journal = "Nucl. Phys. A",
    volume = "750",
    pages = "294--323",
    year = "2005",
    note = "[Erratum: Nucl.Phys.A 780, 90--90 (2006)]"
}

@article{Xu:2015bpl,
    author = "Xu, Si-Qi and Xie, Ju-Jun and Chen, Xu-Rong and Jia, Duo-Jie",
    title = "{The $\Xi^* \bar{K}$ and $\Omega \eta$ interaction within a chiral unitary approach}",
    eprint = "1510.07419",
    archivePrefix = "arXiv",
    primaryClass = "nucl-th",
    doi = "10.1088/0253-6102/65/1/53",
    journal = "Commun. Theor. Phys.",
    volume = "65",
    number = "1",
    pages = "53--56",
    year = "2016"
}

@article{Wang:2008zzz,
    author = "Wang, W. L. and Huang, F. and Zhang, Z. Y. and Liu, F.",
    title = "{Xi anti-K interaction in a chiral model}",
    doi = "10.1088/0954-3899/35/8/085003",
    journal = "J. Phys. G",
    volume = "35",
    pages = "085003",
    year = "2008"
}

@article{Feijoo:2024qgq,
    author = "Feijoo, Albert and Vida{\~n}a, Isaac",
    title = "{On the possible existence of a $S=-3$, $I=1$ pentaquark}",
    eprint = "2411.18248",
    archivePrefix = "arXiv",
    primaryClass = "hep-ph",
    doi = "10.1140/epja/s10050-025-01664-9",
    journal = "Eur. Phys. J. A",
    volume = "61",
    number = "8",
    pages = "196",
    year = "2025"
}

@article{Gamermann:2011mq,
    author = "Gamermann, D. and Garcia-Recio, C. and Nieves, J. and Salcedo, L. L.",
    title = "{Odd Parity Light Baryon Resonances}",
    eprint = "1104.2737",
    archivePrefix = "arXiv",
    primaryClass = "hep-ph",
    doi = "10.1103/PhysRevD.84.056017",
    journal = "Phys. Rev. D",
    volume = "84",
    pages = "056017",
    year = "2011"
}

@article{Sarkar:2010saz,
    author = "Sarkar, Sourav and Sun, Bao-Xi and Oset, E. and Vicente Vacas, M. J.",
    title = "{Dynamically generated resonances from the vector octet-baryon decuplet interaction}",
    eprint = "0902.3150",
    archivePrefix = "arXiv",
    primaryClass = "hep-ph",
    doi = "10.1140/epja/i2010-10956-4",
    journal = "Eur. Phys. J. A",
    volume = "44",
    pages = "431--443",
    year = "2010"
}

@article{Yan:2026yrd,
    author = "Yan, Ye and Wu, Yuheng and Tan, Yue and Huang, Qi and Huang, Hongxia and Ping, Jialun",
    title = "{Investigating $Ωϕ$ Interaction and Correlation Functions}",
    eprint = "2601.01338",
    archivePrefix = "arXiv",
    primaryClass = "hep-ph",
    journal = {arXiv preprint},
    month = "1",
    year = "2026"
}

@article{Encarnacion:2024jge,
    author = "Encarnaci{\'o}n, P. and Feijoo, A. and Sarti, V. Mantovani and Ramos, A.",
    title = "{Femtoscopic study of the S=-1 meson-baryon interaction: K-p, {\ensuremath{\pi}}-{\ensuremath{\Lambda}}, and K+{\ensuremath{\Xi}}- correlations}",
    eprint = "2412.20880",
    archivePrefix = "arXiv",
    primaryClass = "hep-ph",
    doi = "10.1103/3ycr-vzmd",
    journal = "Phys. Rev. D",
    volume = "111",
    number = "11",
    pages = "114013",
    year = "2025"
}

@article{Feijoo:2024bvn,
    author = "Feijoo, A. and Korwieser, M. and Fabbietti, L.",
    title = "{Relevance of the coupled channels in the {\ensuremath{\phi}}p and {\ensuremath{\rho}}0p correlation functions}",
    eprint = "2407.01128",
    archivePrefix = "arXiv",
    primaryClass = "hep-ph",
    doi = "10.1103/PhysRevD.111.014009",
    journal = "Phys. Rev. D",
    volume = "111",
    number = "1",
    pages = "014009",
    year = "2025"
}

@article{Sarti:2023wlg,
    author = "Sarti, V. Mantovani and Feijoo, A. and Vida{\~n}a, I. and Ramos, A. and Giacosa, F. and Hyodo, T. and Kamiya, Y.",
    title = "{Constraining the low-energy S=-2 meson-baryon interaction with two-particle correlations}",
    eprint = "2309.08756",
    archivePrefix = "arXiv",
    primaryClass = "hep-ph",
    doi = "10.1103/PhysRevD.110.L011505",
    journal = "Phys. Rev. D",
    volume = "110",
    number = "1",
    pages = "L011505",
    year = "2024"
}

@article{Pavao:2018xub,
    author = "Pavao, R. and Oset, E.",
    title = "{Coupled channels dynamics in the generation of the $\Omega (2012)$ resonance}",
    eprint = "1808.01950",
    archivePrefix = "arXiv",
    primaryClass = "hep-ph",
    doi = "10.1140/epjc/s10052-018-6329-4",
    journal = "Eur. Phys. J. C",
    volume = "78",
    number = "10",
    pages = "857",
    year = "2018"
}

@article{Nagahiro:2008cv,
    author = "Nagahiro, H. and Roca, L. and Hosaka, A. and Oset, E.",
    title = "{Hidden gauge formalism for the radiative decays of axial-vector mesons}",
    eprint = "0809.0943",
    archivePrefix = "arXiv",
    primaryClass = "hep-ph",
    doi = "10.1103/PhysRevD.79.014015",
    journal = "Phys. Rev. D",
    volume = "79",
    pages = "014015",
    year = "2009"
}

@article{Wang:2022aga,
    author = "Wang, Wen-Fei and Feijoo, Albert and Song, Jing and Oset, Eulogio",
    title = "{Molecular {\ensuremath{\Omega}}cc, {\ensuremath{\Omega}}bb, and {\ensuremath{\Omega}}bc states}",
    eprint = "2208.14858",
    archivePrefix = "arXiv",
    primaryClass = "hep-ph",
    doi = "10.1103/PhysRevD.106.116004",
    journal = "Phys. Rev. D",
    volume = "106",
    number = "11",
    pages = "116004",
    year = "2022"
}

@article{Roca:2024nsi,
    author = "Roca, L. and Song, J. and Oset, E.",
    title = "{Molecular pentaquarks with hidden charm and double strangeness}",
    eprint = "2403.08732",
    archivePrefix = "arXiv",
    primaryClass = "hep-ph",
    doi = "10.1103/PhysRevD.109.094005",
    journal = "Phys. Rev. D",
    volume = "109",
    number = "9",
    pages = "094005",
    year = "2024"
}

@article{Debastiani:2017ewu,
    author = "Debastiani, V. R. and Dias, J. M. and Liang, W. H. and Oset, E.",
    title = "{Molecular $\Omega_c$ states generated from coupled meson-baryon channels}",
    eprint = "1710.04231",
    archivePrefix = "arXiv",
    primaryClass = "hep-ph",
    doi = "10.1103/PhysRevD.97.094035",
    journal = "Phys. Rev. D",
    volume = "97",
    number = "9",
    pages = "094035",
    year = "2018"
}

@article{Hyodo:2013nka,
    author = "Hyodo, Tetsuo",
    title = "{Structure and compositeness of hadron resonances}",
    eprint = "1310.1176",
    archivePrefix = "arXiv",
    primaryClass = "hep-ph",
    reportNumber = "YITP-13-109",
    doi = "10.1142/S0217751X13300457",
    journal = "Int. J. Mod. Phys. A",
    volume = "28",
    pages = "1330045",
    year = "2013"
}

@article{Aceti:2014ala,
    author = "Aceti, F. and Dai, L. R. and Geng, L. S. and Oset, E. and Zhang, Y.",
    title = "{Meson-baryon components in the states of the baryon decuplet}",
    eprint = "1301.2554",
    archivePrefix = "arXiv",
    primaryClass = "hep-ph",
    doi = "10.1140/epja/i2014-14057-2",
    journal = "Eur. Phys. J. A",
    volume = "50",
    pages = "57",
    year = "2014"
}

@article{Wang:2022pin,
    author = "Wang, Zheng-Li and Zou, Bing-Song",
    title = "{Two dynamical generated $a_0$ resonances by interactions between vector mesons}",
    eprint = "2203.02899",
    archivePrefix = "arXiv",
    primaryClass = "hep-ph",
    doi = "10.1140/epjc/s10052-022-10460-4",
    journal = "Eur. Phys. J. C",
    volume = "82",
    number = "6",
    pages = "509",
    year = "2022"
}

@article{ParticleDataGroup:2024cfk,
    author = "Navas, S. and others",
    collaboration = "Particle Data Group",
    title = "{Review of particle physics}",
    doi = "10.1103/PhysRevD.110.030001",
    journal = "Phys. Rev. D",
    volume = "110",
    number = "3",
    pages = "030001",
    year = "2024"
}

@article{Molina:2020hde,
    author = "Molina, R. and Oset, E.",
    title = "{Molecular picture for the $X_0(2866)$ as a $D^* \bar{K}^*$ $J^P=0^+$ state and related $1^+,2^+$ states}",
    eprint = "2008.11171",
    archivePrefix = "arXiv",
    primaryClass = "hep-ph",
    doi = "10.1016/j.physletb.2020.135870",
    journal = "Phys. Lett. B",
    volume = "811",
    pages = "135870",
    year = "2020",
    note = "[Erratum: Phys.Lett.B 837, 137645 (2023)]"
}

@article{Dai:2021vgf,
    author = "Dai, L. R. and Molina, R. and Oset, E.",
    title = "{Prediction of new Tcc states of D*D* and Ds*D* molecular nature}",
    eprint = "2110.15270",
    archivePrefix = "arXiv",
    primaryClass = "hep-ph",
    doi = "10.1103/PhysRevD.105.016029",
    journal = "Phys. Rev. D",
    volume = "105",
    number = "1",
    pages = "016029",
    year = "2022",
    note = "[Erratum: Phys.Rev.D 106, 099902 (2022)]"
}

@article{Dai:2022ulk,
    author = "Dai, L. R. and Oset, E. and Feijoo, A. and Molina, R. and Roca, L. and Torres, A. Mart{\'\i}nez and Khemchandani, K. P.",
    title = "{Masses and widths of the exotic molecular B(s)(*)B(s)(*) states}",
    eprint = "2201.04840",
    archivePrefix = "arXiv",
    primaryClass = "hep-ph",
    doi = "10.1103/PhysRevD.105.074017",
    journal = "Phys. Rev. D",
    volume = "105",
    number = "7",
    pages = "074017",
    year = "2022",
    note = "[Erratum: Phys.Rev.D 106, 099904 (2022)]"
}

@article{Ikeno:2021mcb,
    author = "Ikeno, Natsumi and Molina, Raquel and Oset, Eulogio",
    title = "{$Z_{cs}$ states from the $D^*_s \overline{D}$ and $J/\Psi K^*$ coupled channels: Signal in $B^+ \rightarrow J/\Psi \tau K^*$ decay}",
    eprint = "2111.05024",
    archivePrefix = "arXiv",
    primaryClass = "hep-ph",
    doi = "10.1103/PhysRevD.105.014012",
    journal = "Phys. Rev. D",
    volume = "105",
    number = "1",
    pages = "014012",
    year = "2022",
    note = "[Erratum: Phys.Rev.D 106, 099905 (2022)]"
}

@article{Molina:2008jw,
    author = "Molina, R. and Nicmorus, D. and Oset, E.",
    title = "{The rho rho interaction in the hidden gauge formalism and the f(0)(1370) and f(2)(1270) resonances}",
    eprint = "0809.2233",
    archivePrefix = "arXiv",
    primaryClass = "hep-ph",
    reportNumber = "FTUV-12-0908, IFIC-12-0908",
    doi = "10.1103/PhysRevD.78.114018",
    journal = "Phys. Rev. D",
    volume = "78",
    pages = "114018",
    year = "2008"
}

@article{Yang:2024nss,
    author = "Yang, Zi-Ying and Song, Jing and Liang, Wei-Hong and Oset, Eulogio",
    title = "{$P_c(4440)$ and $P_c(4457)$ decay into $\bar{D}\Sigma _c$ and $\bar{D}\Lambda _c$ and the spin of the $P_c$ states}",
    eprint = "2412.15731",
    archivePrefix = "arXiv",
    primaryClass = "hep-ph",
    doi = "10.1140/epjc/s10052-025-14639-3",
    journal = "Eur. Phys. J. C",
    volume = "85",
    number = "9",
    pages = "954",
    year = "2025"
}

@article{Bramon:1992kr,
    author = "Bramon, A. and Grau, A. and Pancheri, G.",
    title = "{Intermediate vector meson contributions to V0 ---{\ensuremath{>}} P0 P0 gamma decays}",
    reportNumber = "LNF-92-011-P",
    doi = "10.1016/0370-2693(92)90041-2",
    journal = "Phys. Lett. B",
    volume = "283",
    pages = "416--420",
    year = "1992"
}

@article{Geng:2008gx,
    author = "Geng, L. S. and Oset, E.",
    title = "{Vector meson-vector meson interaction in a hidden gauge unitary approach}",
    eprint = "0812.1199",
    archivePrefix = "arXiv",
    primaryClass = "hep-ph",
    doi = "10.1103/PhysRevD.79.074009",
    journal = "Phys. Rev. D",
    volume = "79",
    pages = "074009",
    year = "2009"
}

@article{Oset:2002sh,
    author = "Oset, E. and Pelaez, J. R. and Roca, L.",
    title = "{eta ---{\ensuremath{>}} pi0 gamma gamma decay within a chiral unitary approach}",
    eprint = "hep-ph/0210282",
    archivePrefix = "arXiv",
    reportNumber = "FTUV-02-1021, IFIC-02-1021",
    doi = "10.1103/PhysRevD.67.073013",
    journal = "Phys. Rev. D",
    volume = "67",
    pages = "073013",
    year = "2003"
}

@article{Garzon:2012np,
    author = "Garzon, E. J. and Oset, E.",
    title = "{Effects of pseudoscalar-baryon channels in the dynamically generated vector-baryon resonances}",
    eprint = "1201.3756",
    archivePrefix = "arXiv",
    primaryClass = "hep-ph",
    doi = "10.1140/epja/i2012-12005-x",
    journal = "Eur. Phys. J. A",
    volume = "48",
    pages = "5",
    year = "2012"
}

@book{Ericson:1988gk,
    author = "Ericson, Torleif Erik Oskar and Weise, W.",
    title = "{Pions and Nuclei}",
    isbn = "978-0-19-852008-5",
    publisher = "Clarendon Press",
    address = "Oxford, UK",
    year = "1988"
}

@article{Gamermann:2009uq,
    author = "Gamermann, D. and Nieves, J. and Oset, E. and Ruiz Arriola, E.",
    title = "{Couplings in coupled channels versus wave functions: application to the X(3872) resonance}",
    eprint = "0911.4407",
    archivePrefix = "arXiv",
    primaryClass = "hep-ph",
    doi = "10.1103/PhysRevD.81.014029",
    journal = "Phys. Rev. D",
    volume = "81",
    pages = "014029",
    year = "2010"
}

@article{Albaladejo:2023wmv,
    author = "Albaladejo, M. and Feijoo, A. and Vida{\~n}a, I. and Nieves, J. and Oset, E.",
    title = "{Inverse problem in femtoscopic correlation functions: the $T_{cc}(3875)^+$ state}",
    eprint = "2307.09873",
    archivePrefix = "arXiv",
    primaryClass = "hep-ph",
    doi = "10.1140/epja/s10050-025-01650-1",
    journal = "Eur. Phys. J. A",
    volume = "61",
    number = "8",
    pages = "187",
    year = "2025"
}

@article{Feijoo:2023sfe,
    author = "Feijoo, A. and Dai, L. R. and Abreu, L. M. and Oset, E.",
    title = "{Correlation function for the Tbb state: Determination of the binding, scattering lengths, effective ranges, and molecular probabilities}",
    eprint = "2309.00444",
    archivePrefix = "arXiv",
    primaryClass = "hep-ph",
    doi = "10.1103/PhysRevD.109.016014",
    journal = "Phys. Rev. D",
    volume = "109",
    number = "1",
    pages = "016014",
    year = "2024"
}

@article{ALICE:2021cpv,
    author = "Acharya, Shreyasi and others",
    collaboration = "ALICE",
    title = "{Experimental Evidence for an Attractive p-$\phi$ Interaction}",
    eprint = "2105.05578",
    archivePrefix = "arXiv",
    primaryClass = "nucl-ex",
    reportNumber = "CERN-EP-2021-081",
    doi = "10.1103/PhysRevLett.127.172301",
    journal = "Phys. Rev. Lett.",
    volume = "127",
    number = "17",
    pages = "172301",
    year = "2021"
}

@article{ALICE:2025flv,
    author = "Abualrob, Ibrahim Jaser and others",
    collaboration = "ALICE",
    title = "{First direct access to the $ρ^0$p interaction via correlation studies at the LHC}",
    eprint = "2508.09867",
    archivePrefix = "arXiv",
    primaryClass = "nucl-ex",
    reportNumber = "CERN-EP-2025-179",
    journal = {arXiv preprint},
    month = "8",
    year = "2025"
}

@article{ALICE:2023wjz,
    author = "Acharya, Shreyasi and others",
    collaboration = "ALICE",
    title = "{Accessing the strong interaction between {\ensuremath{\Lambda}} baryons and charged kaons with the femtoscopy technique at the LHC}",
    eprint = "2305.19093",
    archivePrefix = "arXiv",
    primaryClass = "nucl-ex",
    reportNumber = "CERN-EP-2023-106",
    doi = "10.1016/j.physletb.2023.138145",
    journal = "Phys. Lett. B",
    volume = "845",
    pages = "138145",
    year = "2023"
}

@article{ALICE:2022yyh,
    author = "Acharya, Shreyasi and others",
    collaboration = "ALICE",
    title = "{Constraining the ${\overline{\textrm{K}}}{\textrm{N}}$ coupled channel dynamics using femtoscopic correlations at the LHC}",
    eprint = "2205.15176",
    archivePrefix = "arXiv",
    primaryClass = "nucl-ex",
    reportNumber = "CERN-EP-2022-107",
    doi = "10.1140/epjc/s10052-023-11476-0",
    journal = "Eur. Phys. J. C",
    volume = "83",
    number = "4",
    pages = "340",
    year = "2023"
}

@article{Bando:1984ej,
    author = "Bando, M. and Kugo, T. and Uehara, S. and Yamawaki, K. and Yanagida, T.",
    title = "{Is rho Meson a Dynamical Gauge Boson of Hidden Local Symmetry?}",
    reportNumber = "RRK 84-22",
    doi = "10.1103/PhysRevLett.54.1215",
    journal = "Phys. Rev. Lett.",
    volume = "54",
    pages = "1215",
    year = "1985"
}

@article{Bando:1987br,
    author = "Bando, Masako and Kugo, Taichiro and Yamawaki, Koichi",
    title = "{Nonlinear Realization and Hidden Local Symmetries}",
    reportNumber = "DPNU-87-63, AICHI-1, KUNS-903",
    doi = "10.1016/0370-1573(88)90019-1",
    journal = "Phys. Rept.",
    volume = "164",
    pages = "217--314",
    year = "1988"
}

@article{Meissner:1987ge,
    author = "Meissner, Ulf G.",
    title = "{Low-Energy Hadron Physics from Effective Chiral Lagrangians with Vector Mesons}",
    reportNumber = "MIT-CTP-1471",
    doi = "10.1016/0370-1573(88)90090-7",
    journal = "Phys. Rept.",
    volume = "161",
    pages = "213",
    year = "1988"
}

@book{Close1979,
  title     = {An Introduction to Quarks and Partons},
  author    = {Close, F. E.},
  year      = {1979},
  publisher = {Academic Press},
  address   = {London}
}

@book{MandlShaw2010,
  title     = {Quantum Field Theory},
  author    = {Mandl, F. and Shaw, G.},
  publisher = {John Wiley \& Sons},
  address   = {New York},
  year      = {2010},
  note      = {Originally published in 1984}
}

@book{Rose1957,
  title     = {Elementary Theory of Angular Momentum},
  author    = {Rose, M. E.},
  publisher = {John Wiley \& Sons},
  address   = {New York},
  year      = {1957}
}

@article{Song:2022yvz,
    author = "Song, Jing and Dai, L. R. and Oset, E.",
    title = "{How much is the compositeness of a bound state constrained by a and $r_0$? The role of the interaction range}",
    eprint = "2201.04414",
    archivePrefix = "arXiv",
    primaryClass = "hep-ph",
    doi = "10.1140/epja/s10050-022-00753-3",
    journal = "Eur. Phys. J. A",
    volume = "58",
    number = "7",
    pages = "133",
    year = "2022"
}

@article{Albaladejo:2025lhn,
    author = "Albaladejo, Miguel and Canoa, Alejandro and Nieves, Juan and Pel{\'a}ez, Jose Ram{\'o}n and Ruiz-Arriola, Enrique and de Elvira, Jacobo Ruiz",
    title = "{The role of chiral symmetry and the non-ordinary {\ensuremath{\kappa}}/K0{\textasteriskcentered}(700) nature in {\ensuremath{\pi}}{\ensuremath{\pm}}KS femtoscopic correlations}",
    eprint = "2503.19746",
    archivePrefix = "arXiv",
    primaryClass = "hep-ph",
    reportNumber = "IPARCOS-UCM-25-020",
    doi = "10.1016/j.physletb.2025.139552",
    journal = "Phys. Lett. B",
    volume = "866",
    pages = "139552",
    year = "2025"
}

@article{ALICE:2023eyl,
    author = "Acharya, Shreyasi and others",
    collaboration = "ALICE",
    title = "{Investigating the composition of the K0{\textasteriskcentered}(700) state with {\ensuremath{\pi}}{\ensuremath{\pm}}KS0 correlations at the LHC}",
    eprint = "2312.12830",
    archivePrefix = "arXiv",
    primaryClass = "hep-ex",
    reportNumber = "CERN-EP-2023-287",
    doi = "10.1016/j.physletb.2024.138915",
    journal = "Phys. Lett. B",
    volume = "856",
    pages = "138915",
    year = "2024"
}

@article{Sakai:2017hpg,
    author = "Sakai, S. and Oset, E. and Ramos, A.",
    title = "{Triangle singularities in $B^-\rightarrow K^-\pi^-D_{s0}^+$ and $B^-\rightarrow K^-\pi^-D_{s1}^+$}",
    eprint = "1705.03694",
    archivePrefix = "arXiv",
    primaryClass = "hep-ph",
    doi = "10.1140/epja/i2018-12450-5",
    journal = "Eur. Phys. J. A",
    volume = "54",
    number = "1",
    pages = "10",
    year = "2018"
}

\end{document}